\documentclass[a4paper,12pt]{article}
\usepackage{graphicx}
\usepackage{amsmath}
\usepackage{multirow}
\usepackage{float} 

\begin{document}

\begin{flushright}
preprint SHEP-11-13\\
\today
\end{flushright}
\vspace*{1.0truecm}

\begin{center}
{\large\bf Reinforcing the no-lose theorem\\[0.15cm]
for NMSSM Higgs discovery at the LHC}\\
\vspace*{1.0truecm}
{\large M. M. Almarashi and S. Moretti}\\
\vspace*{0.5truecm}
{\it School of Physics \& Astronomy, \\
 University of Southampton, Southampton, SO17 1BJ, UK}
\end{center}

\vspace*{1.0truecm}
\begin{center}
\begin{abstract}
\noindent 
We show the potential of the LHC to detect a CP-even Higgs boson of the NMSSM, $h_1$ or $h_2$, 
decaying into two rather light CP-odd Higgs bosons, $a_1$, by exploiting the production mode based on
Higgs-strahlung off $b$-quarks, i.e., the channel $pp\to b\bar b h_{1,2}$.
We also consider the case of $h_2\to h_1 h_1$ decays.  It is found that these decays have dominant BRs 
over large regions of the NMSSM parameter space where tan$\beta$ is large, a condition 
which also favours the $pp\to b\bar b h_{1,2}$ production rates. Further decays of the
light Higgs boson pairs ($a_1$ and $h_1$) into photon, muon, tau and $b$ final states are also 
considered. The overall production and decay rates
for these processes at inclusive level are sizable and should help
extracting at least one Higgs boson signal over the NMSSM parameter space. 
\end{abstract}
\end{center}

\section{Introduction}

The mechanism responsible for Electro-Weak Symmetry Breaking (EWSB) is still unknown. In the Standard Model (SM)
and its extensions based on Supersymmetry (SUSY), such as the Minimal Supersymmetric Standard Model (MSSM) and
the Next-to-Minimal Supersymmetric Standard Model (NMSSM), the Higgs mechanism is introduced for this primary purpose.
 Such a mechanism predicts the existence of at least one physical Higgs boson, which is a spin zero particle emerging
from EWSB. While only one Higgs boson exists in the SM and five Higgs bosons in the MSSM,
there are seven Higgs bosons in the NMSSM: three CP-even Higgses $h_{1, 2, 3}$ ($m_{h_1} < m_{h_2} < m_{h_3}$), 
two CP-odd Higgses $a_{1, 2}$ ($m_{a_1} < m_{a_2} $) and two charged Higgses \cite{review}. So, the latter 
has a phenomenologically richer Higgs sector than the two former scenarios.

The NMSSM has two additional merits over the MSSM. On the one hand, 
 it can solve the so-called $\mu$-problem of the MSSM \cite{Kim:1983dt} in a natural way by introducing
a new gauge singlet field \cite{Ellis:1988er}. On the other hand, it can relieve the little hierarchy problem 
\cite{BasteroGil:2000bw,Dermisek:2005ar} since a SM-like scalar Higgs boson with mass less than the SM-like Higgs mass LEP limit
is still quite naturally possible over some regions of the 
NMSSM parameter space. In fact, currently, the NMSSM can also explain a possible LEP excess and is definitely 
preferred by EW global fits. This happens when a SM-like Higgs boson of the NMSSM 
can unconventionally decay into two $a_{1}$'s with $m_{a_1}<2m_b$ \cite{excess} (yet notice that this mass
region is highly constrained by ALEPH \cite{Schael:2010aw}
and BaBar \cite{Aubert:2009cka} data). In fact, there is also another possibility in the NMSSM, due to the fact that 
BR($a_1\to \gamma\gamma$) can be dominant and, as a result, BR($a_1\to b\bar b$) is suppressed even though $m_{a_1}>2m_b$ 
\cite{Almarashi:2010jm,Almarashi:2011bf}. Finally,  a CP-even Higgs state ($h_1$ or $h_2$) can naturally 
have a reduced couplings to the $Z$ boson due to the mixing between the singlet and doublet Higgs fields,
making it a natural possibility that the CP-even Higgs state of the NMSSM could have a mass less 
than the LEP limit on a SM-like Higgs mass.

Probing the Higgs sector of the NMSSM is experimentally challenging, as it is not certain that we can
 always detect its physical states.  There has been some work dedicated to explore the detectability of at least
one Higgs boson of the NMSSM at the Large Hadron Collider (LHC)\footnote{Hereafter, we consider 14 TeV as LHC energy.}
and the Tevatron. In particular, some efforts have been made to extend  the so-called 
`no-lose theorem' of the MSSM -- stating that at least one Higgs boson of the MSSM will be discovered at the LHC
via the usual SM-like production
and decay channels throughout the entire MSSM parameter space \cite{NoLoseMSSM} -- to the case of the NMSSM 
\cite{Almarashi:2010jm,Almarashi:2011bf,Ellwanger:2005uu,Ellwanger:2003jt,NoLoseNMSSM1,Shobig2,Almarashi:2011hj}.
By assuming that Higgs-to-Higgs decays are kinematically not allowed, it was realised that
at least one Higgs boson of the NMSSM will be discovered at the LHC. However, this theorem could be violated
if Higgs-to-Higgs and/or 
Higgs-to-SUSY particle (e.g., into neutralino pairs, yielding invisible Higgs signals) decays 
are kinematically accessible \cite{Cyril,NMSSM-Benchmarks}.

 So far, there is no conclusive evidence
that the no-lose theorem can be confirmed in the context of the NMSSM. In order to establish the theorem for this SUSY scenario, 
Higgs-to-Higgs decays should definitely be taken into account though, in particular $h_1\to a_1a_1$. Such a decay 
can in fact be dominant in large regions of the NMSSM parameter space, for instance, for small $A_k$ \cite{Almarashi:2010jm}, 
and may not give Higgs signals with sufficient significance at the LHC. 
The importance of Higgs-to-Higgs decays in the context of the NMSSM has been emphasised over the years in much literature
in all above respects, see, e.g., Refs.~\cite{Dermisek:2005ar,Gunion:1996fb,Dobrescu:2000jt,Dobrescu:2000yn}.
 Eventually, it was realised that Vector Boson Fusion (VBF)\footnote{Which is dominated by $W^+W^-$-fusion over $ZZ$-one. }
 could be a viable production channel to detect $h_{1,2}\to a_1a_1$ at the LHC,
 in which the Higgs pair decays into $jj\tau^+\tau^-$ \cite{Ellwanger:2005uu,Ellwanger:2003jt,Ellwanger:2004gz}.
 Some scope could also be afforded by a $4\tau$ signature in both VBF and Higgs-strahlung 
(off gauge bosons) \cite{Belyaev:2008gj}. The gluon-fusion channel too could be a means of
 accessing $h_1\to a_1a_1$ decays, so long that the light CP-odd Higgs states both decay 
into four muons \cite{Belyaev:2010ka} or two muons and two taus
\cite{2mu2tau}. Such results were all supported by simulations based on parton shower Monte Carlo (MC) programs and
some level of detector response.

In this paper, we want to investigate whether or not the no-lose theorem of
 the NMSSM at the LHC can possibly be reinforced by considering a Higgs production channel so
 far neglected, i.e., Higgs boson production in association with $b$-quark pairs 
(aka Higgs-strahlung off $b$-quark pairs). Notice that the twin process in 
which $b$-quarks are replaced by $t$-quarks was discussed in \cite{Shobig2},
 where it was found to be very subleading over the NMSSM parameter space. 
We will be looking at inclusive event rates in presence of  
various Higgs-to-Higgs decays, $h_{1, 2}\to a_1a_1$ and $h_2\to h_1h_1$, for $h_1$ and $h_2$ produced
in association with $b$-quark pairs. Notice that this production mode is in general the largest one in the NMSSM at large values of 
tan$\beta$. We will also be studying the decay patterns of the lightest Higgs boson pairs, 
$a_1a_1$ or $h_1h_1$, into 
different types of decay modes.

This paper is organised as follows. In Sec. 2, we describe the parameter space scan performed. Inclusive event rates
for the signals are explained in Sec. 3. Sec. 4 discusses possible
signatures. Finally, we summarise and conclude in Sec. 5.\\

\section{\large Parameter Space Scan}
\label{sect:scan}

Due to the large number of parameters in the NMSSM, it is practically not feasible to do a comprehensive scan 
over all of them. Their number can however be reduced significantly by assuming certain conditions of unification.
Since the mechanism of SUSY breaking is still unknown,
to explore the NMSSM Higgs sector, we have performed a general scan in parameter space by fixing the soft 
SUSY breaking terms at high scale to reduce
their contributions to the outputs of the parameter scans. Consequently, we are left with six independent inputs. 
Our parameter space is in particular defined through the Yukawa couplings $\lambda$ and
$\kappa$, the soft trilinear terms $A_\lambda$ and $A_\kappa$ plus 
tan$\beta$ (the ratio of the 
Vacuum Expectation Values (VEVs) of the two Higgs doublets) and $\mu_{\rm eff} = \lambda\langle S\rangle$
(where $\langle S\rangle$ is the vacuum expectation value of the Higgs singlet). 

We used here the fortran package NMSSMTools developed in Refs.~\cite{NMHDECAY,NMSSMTools}. This package
computes the masses, couplings and decay widths of all the Higgs
bosons of the NMSSM, including radiative corrections, in terms of its parameters at the EW
scale. NMSSMTools also takes into account theoretical as well as
experimental constraints from negative Higgs searches at LEP \cite{LEP} and the Tevatron
as well as other contexts ($B$-physics, low energy experiments, etc.), 
including the unconventional channels relevant for the NMSSM.
 
 We have used the NMHDECAY code to scan over the six tree level parameters of the NMSSM Higgs sector
in the following intervals:
\begin{center}
$\lambda$ : 0.0001 -- 0.7,\phantom{aa} $\kappa$ : 0 --
0.65,\phantom{aa} $\tan\beta$ : 1.6 -- 54,\\ $\mu$ : 100 -- 1000 GeV,\phantom{aa} 
$A_{\lambda}$ : $-$1000 -- +1000 GeV,\phantom{aa} $A_{\kappa}$ :$-$10 -- 0.\\
\end{center}
\noindent
Remaining soft terms, contributing at higher order level, which are fixed in the scan include:\\
$\bullet\phantom{a}m_{Q_3} = m_{U_3} = m_{D_3} = m_{L_3} = m_{E_3} = 1$ TeV, \\
$\bullet\phantom{a}A_{U_3} = A_{D_3} = A_{E_3} = 1.2$ TeV,\\
$\bullet\phantom{a}m_Q = m_U = m_D = m_L = m_E = 1$ TeV,\\
$\bullet\phantom{a} M_1 = M_2 = M_3 = 1.5$ TeV.\\
Notice that the sfermion mass parameters and the $SU(2)$ gaugino mass parameter, $M_2$, play 
crucial roles in constraining tan$\beta$. Decreasing values of those parameters allow smaller values of tan$\beta$ to 
pass experimental and theoretical constraints. In fact, when tan$\beta$ is large the sfermion masses should be large
to avoid the constraints coming from the muon anomalous magnetic moment \cite{Domingo:2008bb}. The dominant supersymmetric contribution
at large tan$\beta$ is due to chargino-sneutrino loop diagram \cite{Czarnecki:2001pv}. Also, notice that the chargino masses
depend strongly on $M_2$. As mentioned above, we fixed the gaugino mass parameters and other SUSY breaking terms at high scale
to reduce their contributions to the outputs of the parameter scans. 

In line with the assumptions made in \cite{Ellwanger:2005uu,Ellwanger:2003jt,NoLoseNMSSM1}, the allowed decay 
modes for neutral NMSSM Higgs bosons are\footnote{Here, we use the label
$h(a)$ to signify any of the neutral CP-even(odd) Higgs bosons of the NMSSM.}:
\begin{eqnarray*}
h,a\rightarrow gg,\phantom{aaa} h,a\rightarrow \mu^+\mu^-,
&&h,a\rightarrow\tau^+\tau^-,\phantom{aaa}h,a\rightarrow
b\bar b,\phantom{aaa}h,a\rightarrow t\bar t, \\ h,a\rightarrow
s\bar s,\phantom{aaa}h,a\rightarrow
c\bar c,&&h\rightarrow W^+W^-,\phantom{aaa}h\rightarrow ZZ, \\
h,a\rightarrow\gamma\gamma,\phantom{aaa}h,a\rightarrow
Z\gamma,&&h,a\rightarrow {\rm Higgses},\phantom{aaa}h,a\rightarrow
{\rm sparticles}.
\end{eqnarray*}

We have performed a random scan over 20 million points in the specified parameter space and required
that $m_{h_2}\leq$ 300 GeV. The output of the scan, as
stated earlier, contains masses, Branching Ratios (BRs) and couplings of
the NMSSM Higgses, for all the successful points, which have passed 
the various experimental and theoretical constraints.

\section{\large Inclusive Event Rates}
\label{sect:rates}
For successful data points, we used CalcHEP \cite{CalcHEP} to determine the
cross-sections for NMSSM Higgs production\footnote{We adopt herein
CTEQ6L \cite{cteq} as parton distribution functions, with scale $Q=\sqrt{\hat{s}}$, the centre-of-mass energy
at parton level, for all processes computed.
Further, we have taken 
$m_b(m_b)=4.214$ GeV for the (running) bottom-quark mass.}.
As the SUSY mass scales have been arbitrarily set well above the EW one (see above), 
the NMSSM Higgs production modes
exploitable in simulations at the LHC are those involving couplings to
heavy ordinary matter only. Amongst the production channels onset by the latter, 
we focus here on the processes 
\begin{equation}
gg, q\bar q\to b\bar b~{h_1} \phantom{aa} {\rm and} \phantom{aa}   gg, q\bar q\to b\bar b~{h_2},
\label{eq:proc}
\end{equation}
i.e., Higgs production in association with a $b$-quark pair. This production mode is dominant at large $\tan\beta$.

To a good approximation, at large $\tan\beta$, the tree level lightest neutral Higgs boson masses are given by
the following expressions \cite{MNZ}: 
\begin{equation*}
 m^2_{a_1}=-\frac{3\kappa\mu_{\rm eff}A_{\kappa}}{\lambda},
\end{equation*}

\begin{eqnarray*}
m^2_{h_{1/2}}&=&\frac{1}{2} \Bigg\{ 
m^2_Z+\frac{\kappa\mu_{\rm eff}}{\lambda}\bigg(\frac{4\kappa\mu_{\rm eff}}{\lambda}+A_{\kappa}\bigg)\\
&& \mp {\sqrt{\Bigg[m^2_Z-\frac{\kappa\mu_{\rm eff}}{\lambda}\bigg(\frac{4\kappa\mu_{\rm eff}}{\lambda}+A_{\kappa}\bigg)\Bigg]^2+
\frac{\lambda^2\upsilon^2}{2{\mu^2_{\rm eff}}}\Bigg[4\mu^2_{\rm eff}-m^2_A\sin^22\beta\Bigg]^2}}\Bigg\},
\end{eqnarray*} 
where 
\begin{equation*}
 m^2_A=\frac{2\mu_{\rm eff}}{\sin2\beta}\bigg(A_{\lambda}+\frac{\kappa\mu_{\rm eff}}{\lambda}\bigg).
\end{equation*}
\noindent
(We reproduced here these tree level formulae mainly for guidance in interpreting the upcoming figures, 
the reader should recall though that NMSSMTools includes radiative corrections as well.)

To probe the NMSSM parameter space, we have computed $m_{h_1}$ and $m_{h_2}$ against each of
the six parameters of the NMSSM (Figs. 1 and 2). As it is clear from the two figures, in our chosen 
parameter space regions, small values of $\lambda$, $\kappa$ and $\mu_{\rm eff}$ are favoured whereas large values
of tan$\beta$ and positive values of $A_{\lambda}$ are the most compatible with theoretical and experimental data.
The distribution over $A_{\kappa}$ is uniform primarily because only small negative values of $\kappa$ are scanned over.

Fig. 3 shows the correlations between all three Higgs masses,
$m_{a_1}$, $m_{h_1}$ and $m_{h_2}$. Since the successful points emerging from the scan have small
values of $\lambda$, $\kappa$ and also $A_{\kappa}$, only rather small values of $m_{a_1}$ are allowed.
It is remarkable that the smaller $m_{a_1}$ the smaller $m_{h_1}$ and $m_{h_2}$ (two top-panes). 
In the bottom-pane of the same figure, for $m_{h_2}$ around 120 GeV, $m_{h_1}$
can have values from just above 0 up to slightly less than 120 GeV, showing the possibility that the two Higgs
states can simultaneously have the same mass, $m_{h_1} \sim m_{h_2}$. 
Notice also that the majority of points have $m_{h_1}$ between 115 GeV and 120 GeV, i.e., just above 
the LEP limit on
a SM-like Higgs mass.

The production times decay rates of $h_1$ and $h_2$, in which $h_1$ decays into two lighter $a_1$'s
and $h_2$ decays into either a pair of $a_1$'s or a pair of $h_1$'s, as functions of the Higgs masses
$m_{h_1}$ and $m_{h_2}$ (left-panes), tan$\beta$ (middle-panes) and of the corresponding Higgs-to-Higgs decays BRs (right-panes), 
are shown in Fig.4. In our choice of parameter space which has large tan$\beta$ we have noticed that the production rate of $h_1$ in 
association with a bottom-antibottom,
$\sigma(pp\to b\bar bh_1)$, is nearly constant, does not depend on the tree level parameters, while
the production rate of $h_2$, $\sigma(pp\to b\bar bh_2)$, is strongly dependent on tan$\beta$ and other tree level parameters.
In fact, notice that in the figure we multiply the production rates by the decay rates of Higgs-to-Higgs particles, 
which play crucial roles in the changes of the inclusive cross section. The two bottom middle-panes of the figure make clear that
while large $\tan\beta$ values are a necessary condition for large production times decay rates of $h_2$ they are not
a sufficient one, as most of the points accumulate at intermediate event rates.

It is clear that Higgs-to-Higgs decays
are dominant over a large area of NMSSM parameter space if Higgs-to-Higgs decays are kinematically allowed
and so these decays should be taken seriously before 
claiming any validity (or otherwise) of the no-lose theorem for the NMSSM, see right-panes of Fig. 4. Fortunately, for considerable 
regions of parameter space, with 
different masses of $h_1$ and $h_2$, these production rates are sizable (up to 1000 fb or so),
except for the case of $h_2\to h_1h_1$ where
only few points have large production rates, due to smallness of BR($h_2\to h_1h_1$) compared with
BR($h_{1, 2}\to a_1a_1$) in general.

Fig. 5 displays the correlations between the three discussed production and decay processes.
 It is quite remarkable that the overall trend, despite an obvious spread also in the horizontal and
vertical directions, is such that when one channel grows in event yield there is also another one which also does,
 hence opening up the possibility of the simultaneous discovery of several Higgs states of the NMSSM 
(other than $a_1$ also $h_1$ {\sl and} $h_2$), an exciting 
prospect in order to distinguish the NMSSM Higgs sector from the MSSM
one.

In Fig. 6 we have calculated the signal rates of $h_1$ and $h_2$ through their cascade decays that
finish with $a_1\to b\bar b$ and/or $a_1\to \tau^+\tau^-$. It is shown that the signal rates are quite large,
topping 1000 fb for $h_1$ and 100 fb for $h_2$
in case of 4$b$ and 4$\tau$ final states due to the fact that BR$(a_1\to b\bar b)$ is dominant
when $m_{a_1}$ $\geq$ 10 GeV and BR$(a_1\to \tau^+\tau^-)$ is dominant for $m_{a_1}$ $<$ 10 GeV.
The 2$b$ plus 2$\tau$ rates are one order of 
magnitude less than the former two due to the fact that only the parameter space points with $m_{a_1}$ $\geq$ 10 GeV
have these final states
in which BR$(a_1\to \tau^+\tau^-)$ is only about 10\% of BR$(a_1\to b\bar b)$. Overall, there are some regions of
parameter space which have considerable signal rates that could be sufficient to  discover the $h_1$ and $h_2$ 
through their $a_1a_1$ cascade decays at the LHC. 

The $h_2$ cascade decays ending with $h_1\to b\bar b$ and/or $h_1\to \tau^+\tau^-$ have less
cross section (see Fig. 7). Only for $m_{h_2}$ around 120 GeV the rates are quite sizable,
topping 50 fb, 5 fb and 0.5 fb level for $4b$, $2b$ plus $2\tau$ and $4\tau$ final states, respectively.

As explained in Ref.~\cite{Almarashi:2010jm}, the BR($a_1\to \gamma\gamma$) can be dominant over a sizable
region of NMSSM parameter space.
This very peculiar phenomenon appears in this SUSY scenario (unlike the MSSM) 
because of the fact that a rather light
CP-odd Higgs state can have a predominant singlet component and
a very weak doublet one. As a consequence, all SM-like partial decay widths are heavily suppressed as
they employ only the doublet component, except one: $\Gamma(a_1\to\gamma\gamma)$. This
comes from the fact that the  $a_1\to \tilde\chi^+\tilde\chi^-$ coupling is not suppressed, as it is generated through
the $\lambda H_1H_2S$ Lagrangian term and therefore implies no small mixing. Although the direct
decay $a_1\to \tilde\chi^+\tilde\chi^-$ is forbidden, the aforementioned coupling participates in 
the $a_1\gamma\gamma$  effective coupling. In hence, when 
BR$(a_1\to\gamma\gamma)$ is very large, no other SM-like BR can be. Hence, 
it makes sense to look at the scope of $a_1a_1\to\gamma\gamma\gamma\gamma$ decays. The corresponding 
inclusive rates are found in Fig. 8 as functions of $m_{h1}$ and $m_{h2}$ (two top-panes) 
for both $h_1\to a_1a_1\to 4\gamma$ and $h_2\to a_1a_1\to 4\gamma$.
Despite inclusive rates are never very large, it should be noticed a consistent population of points in
the former case at $m_{h_1}\approx $ 115 GeV yielding up to ${\cal O}$(1 fb) rates, with also a possibility
of rates reaching up to 100 fb for smaller $m_{h_1}$, and in the latter
case well spread out in $m_{h_2}$ from 115 to 300 GeV yielding some points between 0.1 and 1 fb. 
Moreover, we have shown in the same figure the inclusive results for $h_1\to a_1a_1\to \tau^+\tau^-\mu^+\mu^-$ 
and $h_2\to a_1a_1\to \tau^+\tau^-\mu^+\mu^- $. The rates for $h_1$ can reach 1 fb$^{-1}$ for various 
ranges of $m_{h_1}$ and roughly 0.5 fb$^{-1}$ in case of $h_2$ for essentially any
$m_{h_2}$. 

Finally, notice that the cases $h_1\to a_1a_1\to \mu^+\mu^-\mu^+\mu^-$ and 
$h_2\to a_1a_1\to \mu^+\mu^-\mu^+\mu^-$ contribute below the 0.01 fb level over the 
entire NMSSM parameter space, so we do not show the corresponding plots. 

\section{Possible Signatures}

The production times decay rates presented in the previous section are inclusive results, 
whereby no cuts have been enforced on the final state particles\footnote{Recall that we use 
a finite $b$-quark mass, see Footnote 4.}. Clearly, in order to detect the latter, a typical
 finite volume of an LHC detector has to be emulated. Further, in order
to assess the significance of the signal yield, a background simulation (within the same detector region)
 has to eventually be carried out. Here, in the spirit of Ref.~\cite{Shobig2}, we would like to discuss
 the possible scope of the possible aforementioned signatures, without however
venturing in such a complicated simulations. The key issue to be accessed is clearly
 whether one or more of the $b$-quarks produced in association with the Higgs state $h_1$ or $h_2$
 (henceforth called `prompt' $b$-quarks) in process (\ref{eq:proc}) ought to be tagged as such. 
The relevance of this should be clear from inspecting Fig. 9. The 
$b$-quarks in the final state often emerges from the splitting of a gluon inside the proton, hence they
 can be at very low transverse momentum (denoted here by $p_{T_b}$). To enforce vertex tagging with good 
efficiency, say $\varepsilon_b=60\%$, a minimum $p_{T_b}$ value is always required, the lowest reasonable
 threshold being 15 GeV or so \cite{ATLAS,CMS}. For the case of a single $b$-tag
the overall efficiency at very low Higgs masses (irrespectively of 
considering either $h_1$ or $h_2$ being produced)
is some 2--3\%, eventually growing to 14--15\% for very massive objects. For the case of a double $b$-tag,
we are instead speaking or corresponding rates at the 1\% to 8\% level, respectively. Clearly
then, the scope of the production and decay channels investigated in the previous section much depends
 on the Higgs mass produced and the decay patterns pursued. 

We reckon that for the $4\gamma$ signature it should not be necessary to tag any of the `prompt' $b$-quarks at all, as any of the
(typically high transverse
 momentum and isolated) decay products of the $a_1$'s
 could act as trigger and the SM backgrounds
(which would generally be induced by non-QCD processes) should not be prohibitively large\footnote{This is in fact 
very important in view of the fact that the $4\gamma$ decay rate is the smallest one amongst those studied here.}.
 Regarding signatures with $\tau's$, for $4\tau$ and $2\tau 2b$ one could certainly exploit 
a $\tau$ trigger (both leptonic and hadronic) \cite{ATLAS,CMS}, however (especially in case of hadronic $\tau$ decays),
 it may be necessary to tag at least one `prompt' $b$-quark to suppress QCD backgrounds mimicking $\tau\to$ hadrons. 
The case  $2\tau 2\mu$ would clearly exploit a muon trigger instead. 
Finally, the case of a $4b$ signature of $h_{1,2}\to a_1a_1$ and $h_2\to h_1h_1$
 decays is totally unexplored, especially considering
 the fact that the entire final state would be made up of six quarks, i.e., with an unavoidable huge combinatorics
 and burdened by an extremely large pure QCD background. 

In essence, only a dedicated kinematical analysis of the decay products could in the end ascertain the true 
selection efficiency of a signature and its scope. What we can responsibly do here is to highlight
three possible scenarios. Firstly, one whereby the signal rates in the proceeding section will not be reduced
 substantially after enforcing acceptance cuts: this is certainly applicable to $4\gamma$ events emerging 
from $a_1$ states with masses between 50 and 100 GeV (where the BR$(a_1\to\gamma\gamma)$ is maximal, see
 bottom-left pane of Fig. 2 in \cite{Almarashi:2010jm}) and $4\tau$ events (with the heavy leptons decaying leptonically 
to electron and muons, which however induce a 1\% suppression because of the consequent BRs).
 Secondly, one whereby all decay signatures involving 
(one or more) hadronic $\tau's$ and $b$'s are reduced by a factor between 7 and 50, depending on
the produced Higgs mass, assuming a single tag only of `prompt' $b$'s. Thirdly, one whereby most 
possibly the $6b$ final state requires a double tag of `prompt' $b$-quarks, reducing the signal yield
by a factor between 20 and 100,
depending on the $h_{1,2}$ mass\footnote{Notice that for $a_1$ masses comparable to typical transverse
momentum thresholds of the decay products further severe reductions could occur, however, there is plenty
of NMSSM parameter space giving sizable signals for heavier
$a_1$ states for all signatures considered here.}.  

\section*{Conclusions}
Searching for NMSSM Higgs states at the LHC is very complicated compared to the MSSM ones due to the 
dominance of Higgs-to-Higgs decays in large parameter space regions of the next-to-minimal SUSY
model. This is the main reason why a no-lose theorem
has not been confirmed yet in the context of the NMSSM.
In view of this and following on previous work,
where the case of VBF and Higgs-strahlung of $W,Z$ bosons and
$t$ quarks was studied \cite{Shobig2}, we have found here that, at large values of tan$\beta$, $h_1$ and
$h_2$ production in association with bottom-antibottom pairs and decaying into lighter Higgses can  
have sizable signal rates in some regions of NMSSM parameter space, in a variety of decay patterns including 
photons, muons, tauons and $b$-quarks themselves. We have verified this at the inclusive level and discussed 
what could happen in presence of acceptance cuts and consequent detector efficiencies. 

Clearly, in the end, only a dedicated decay analysis, in presence of
not only acceptance but also selection cuts (the latter driven by the also necessary background assessment),
 will decree whether signal extraction is possible and through which signatures. However,  
our present study, alongside the findings of \cite{Shobig2}, should eventually direct the NMSSM parameter
 space exploration where discovery significances can be found. In all circumstances, just like
with other previous attempts at extracting NMSSM Higgs-to-Higgs signatures, evidence of those investigated
here will require a rather large LHC luminosity sample, of ${\cal O}$(100 fb$^{-1}$) or more.

\section*{Acknowledgments}
This work is 
supported in part by the NExT Institute. M. M. A. acknowledges
a scholarship granted to him by Taibah University (Saudi Arabia).

\begin{figure}
 \centering\begin{tabular}{cc}
  \includegraphics[scale=0.60]{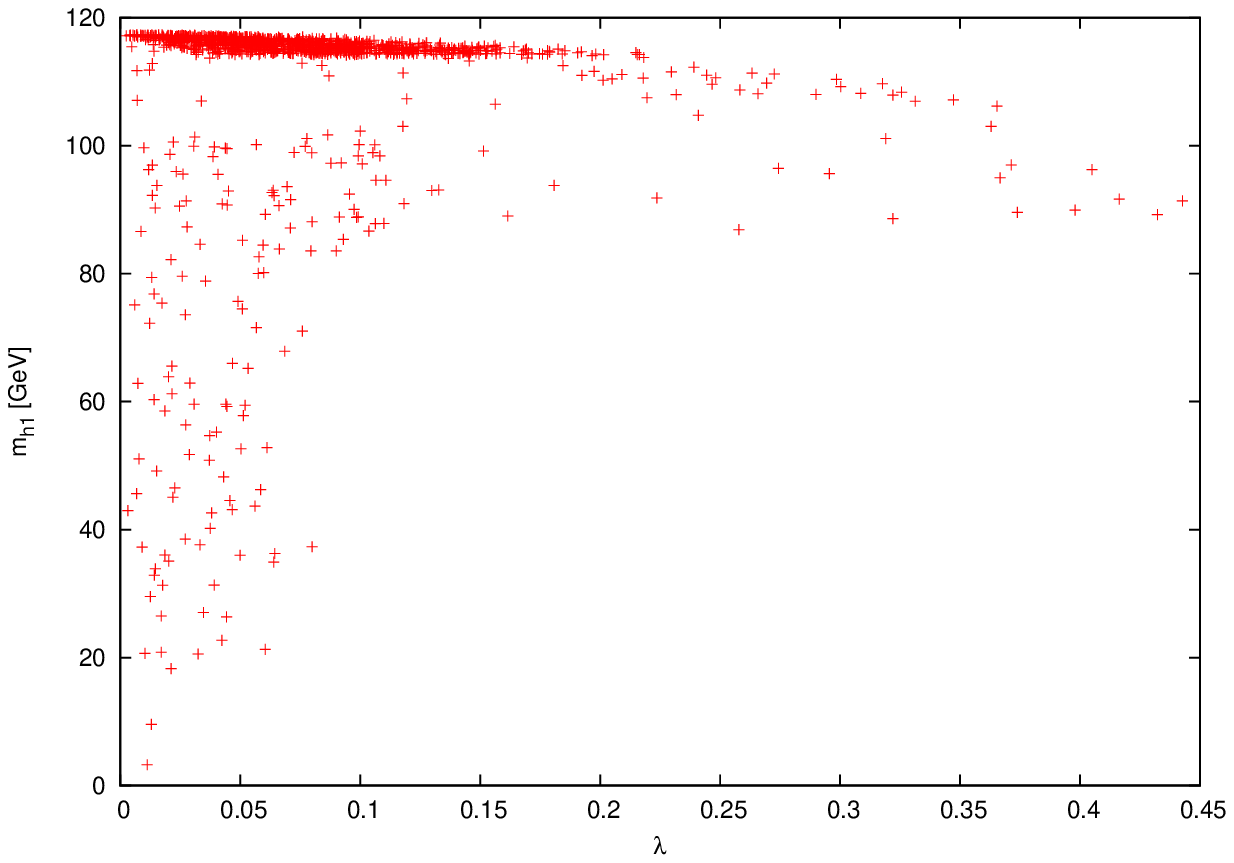}&\includegraphics[scale=0.60]{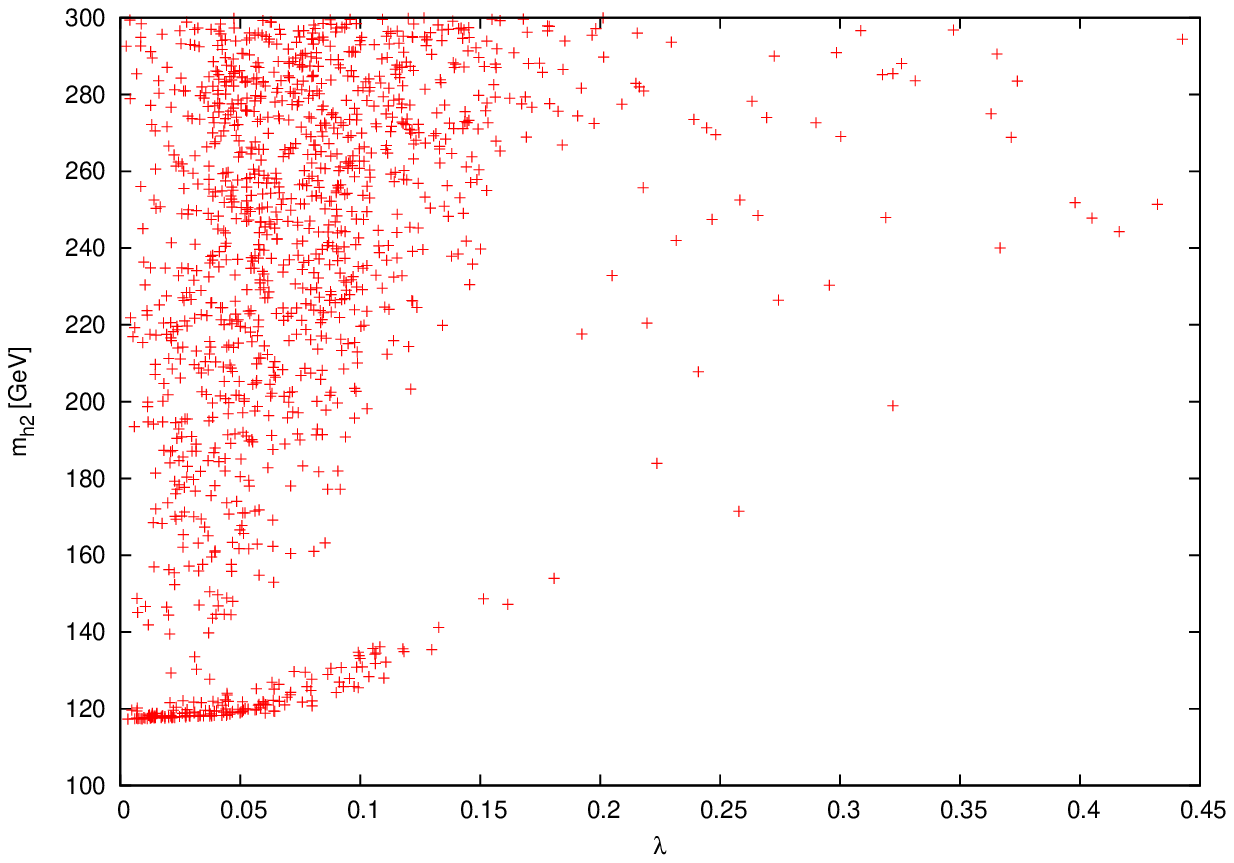}\\
   \includegraphics[scale=0.60]{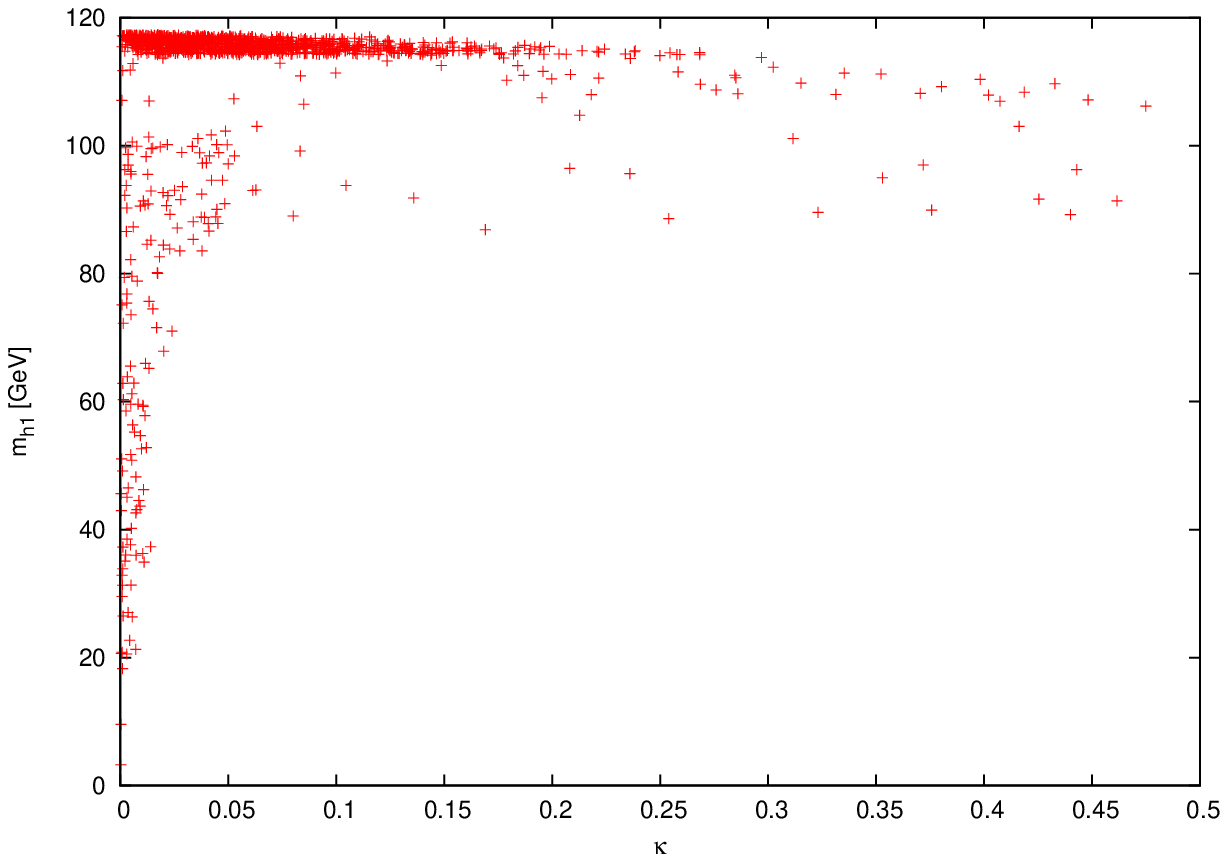}&\includegraphics[scale=0.60]{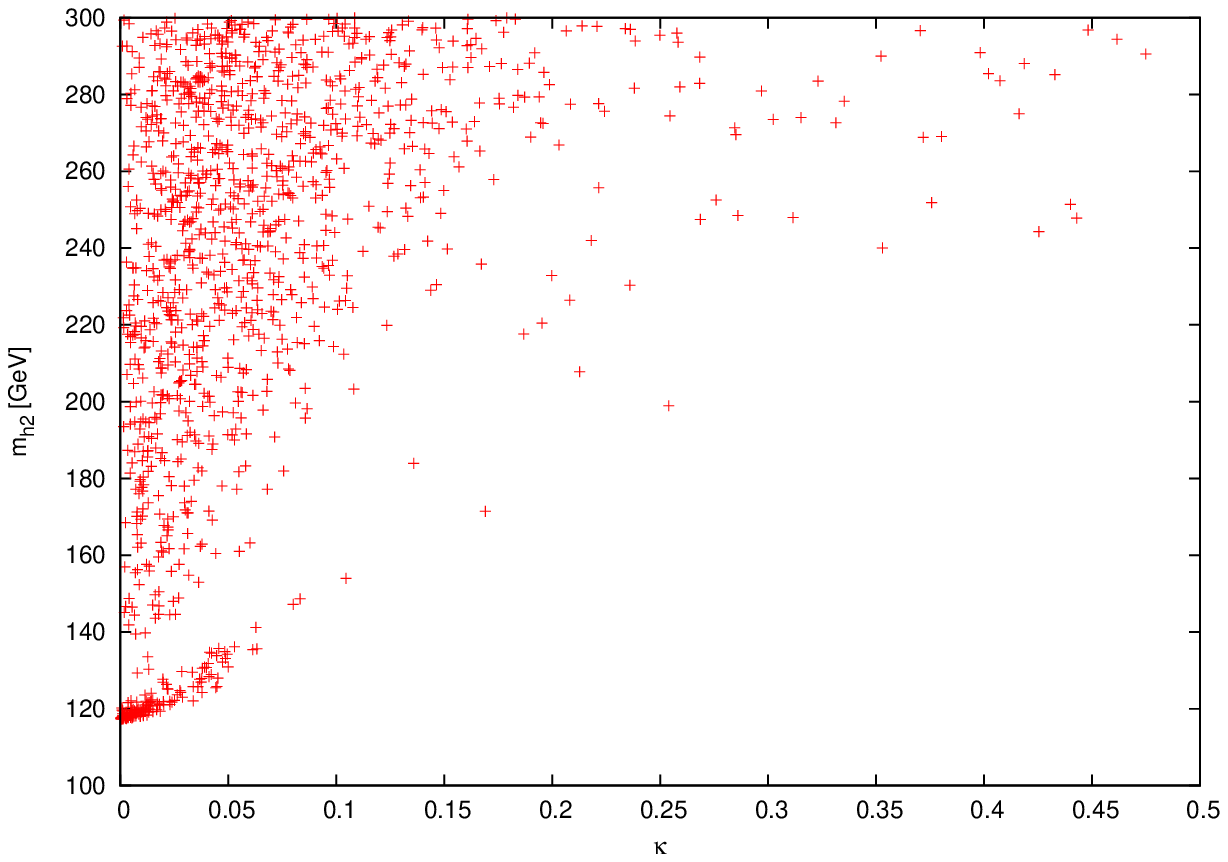}\\
  \includegraphics[scale=0.60]{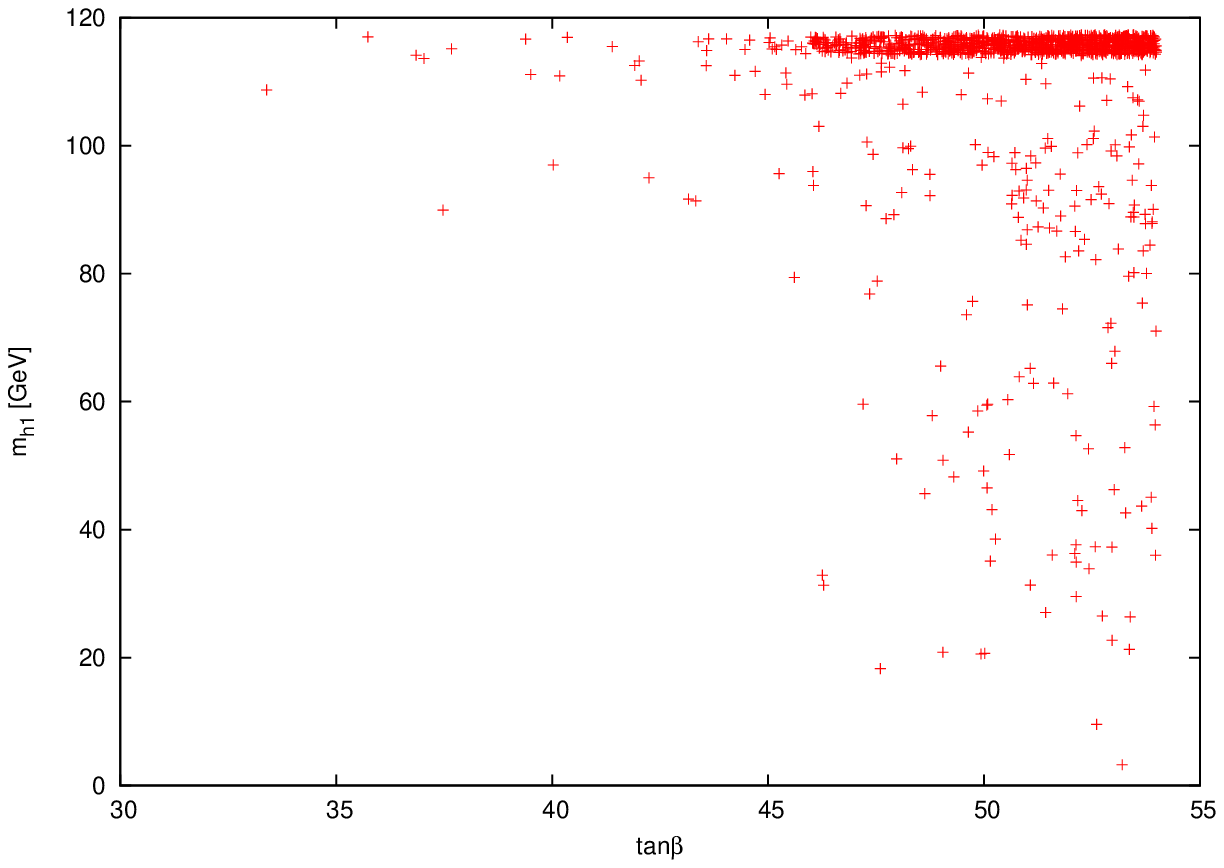}&\includegraphics[scale=0.60]{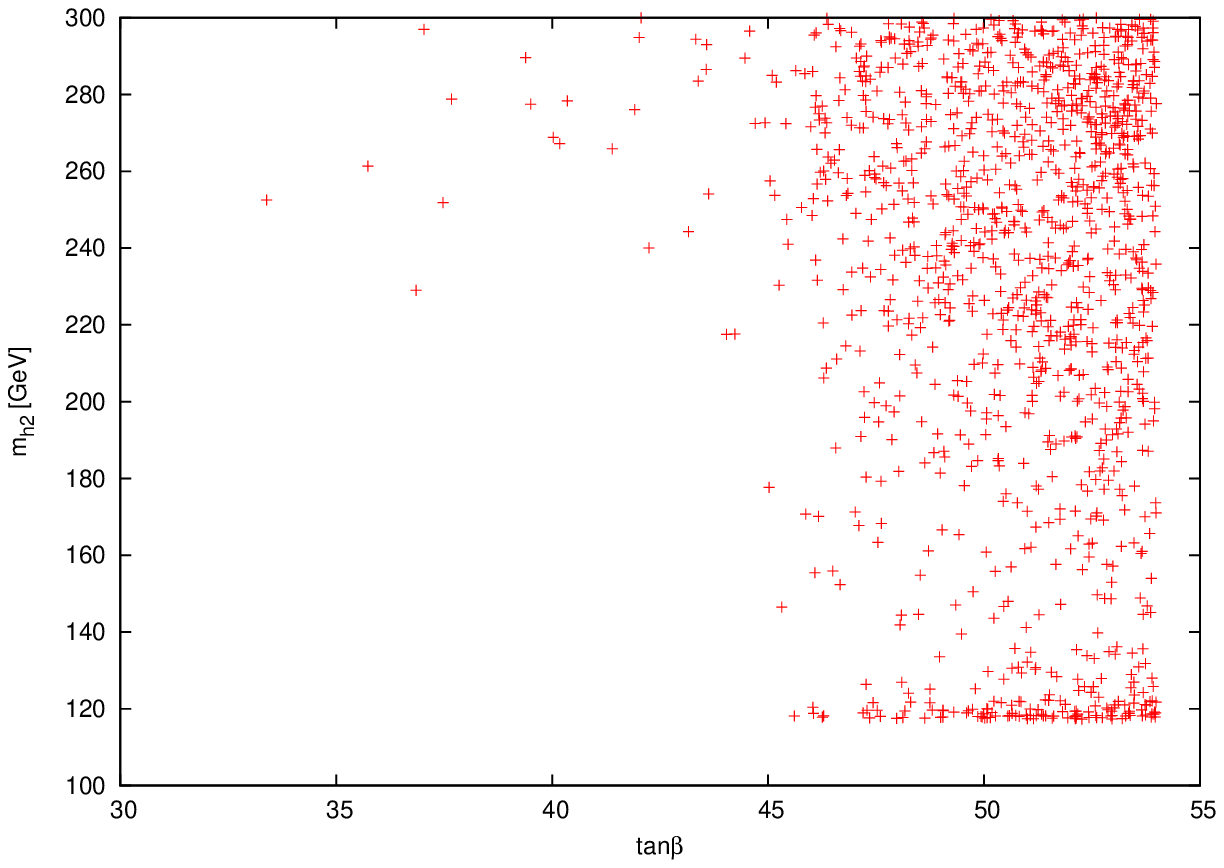}
   \end{tabular}
\label{fig:m1,2-scan1}
\caption{The lightest two scalar Higgs masses $m_{h_1}$ and $m_{h_2}$ as functions of $\lambda$, 
$\kappa$ and $\tan\beta$. }

\end{figure}

\begin{figure}
 \centering\begin{tabular}{cc}
  \includegraphics[scale=0.60]{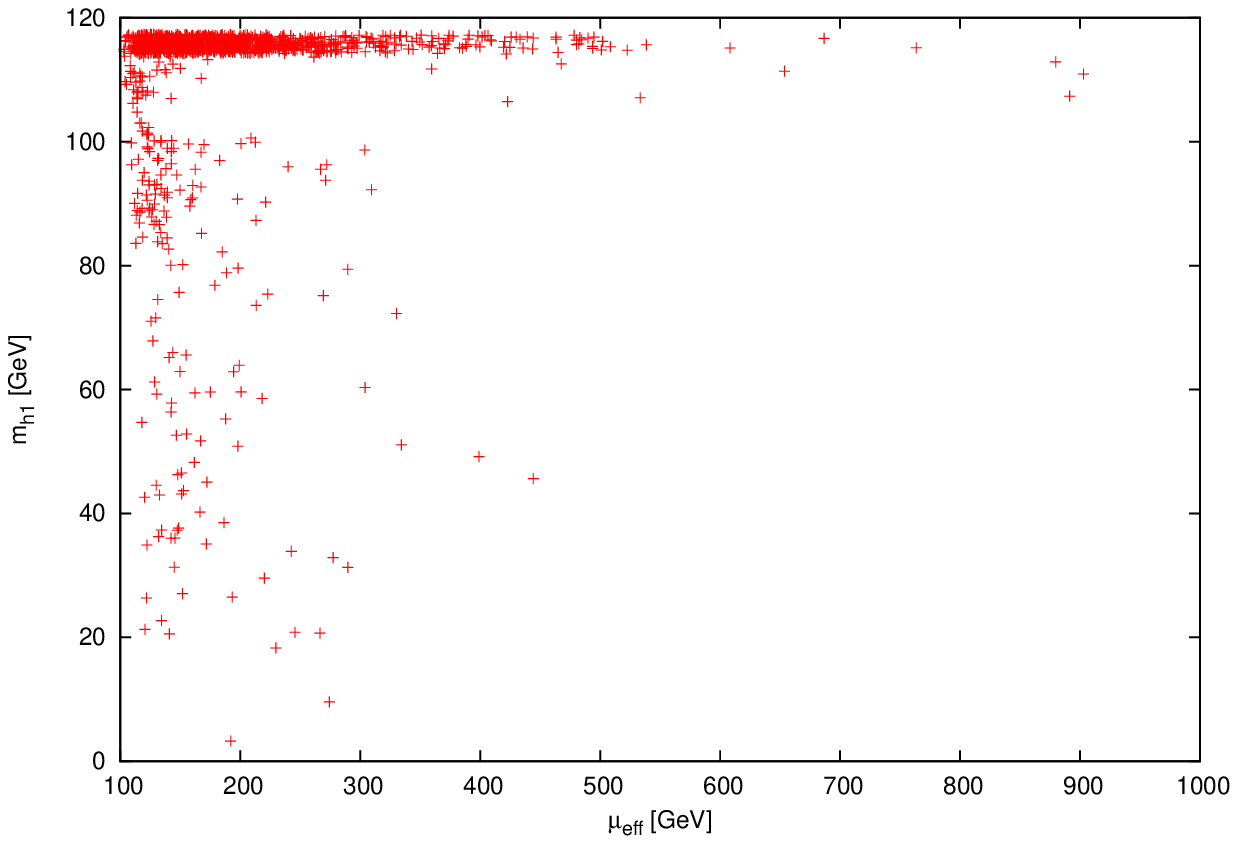}&\includegraphics[scale=0.60]{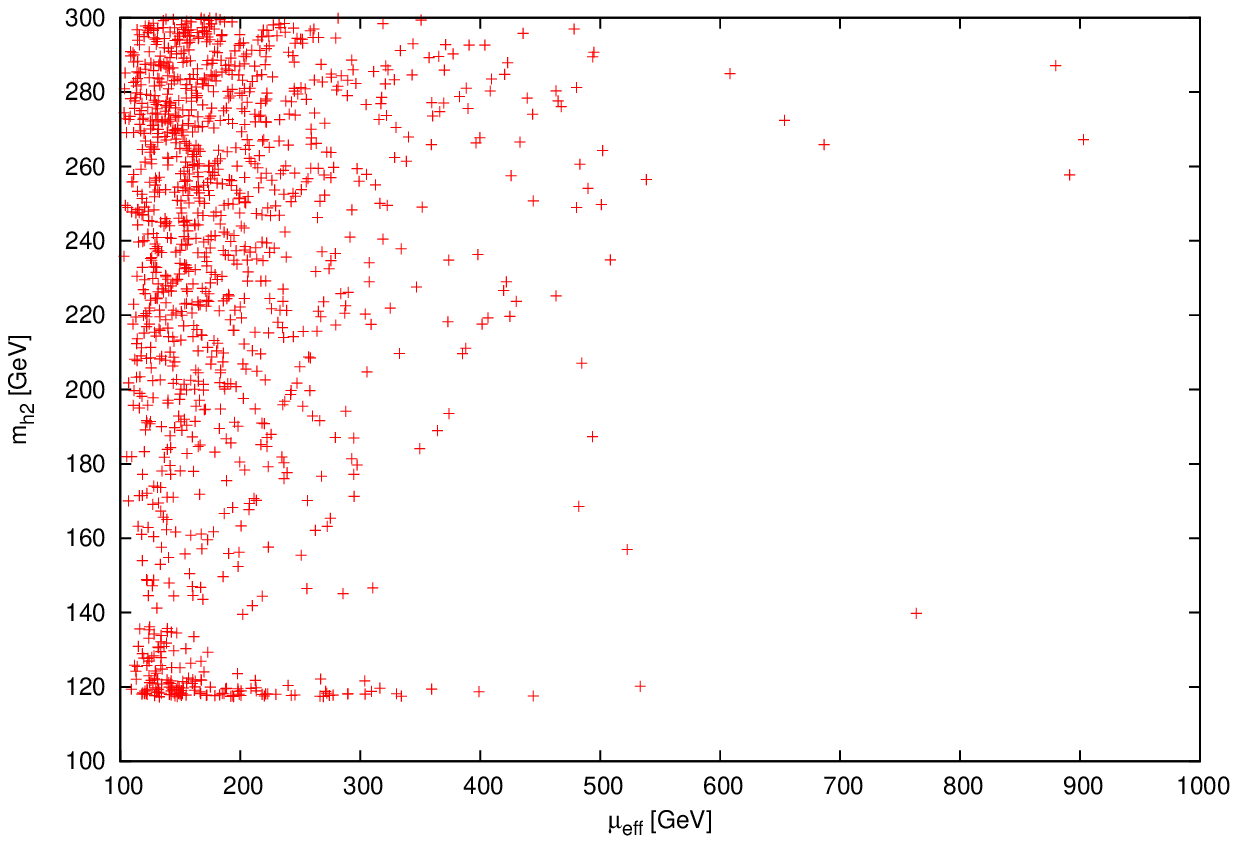}\\
  \includegraphics[scale=0.60]{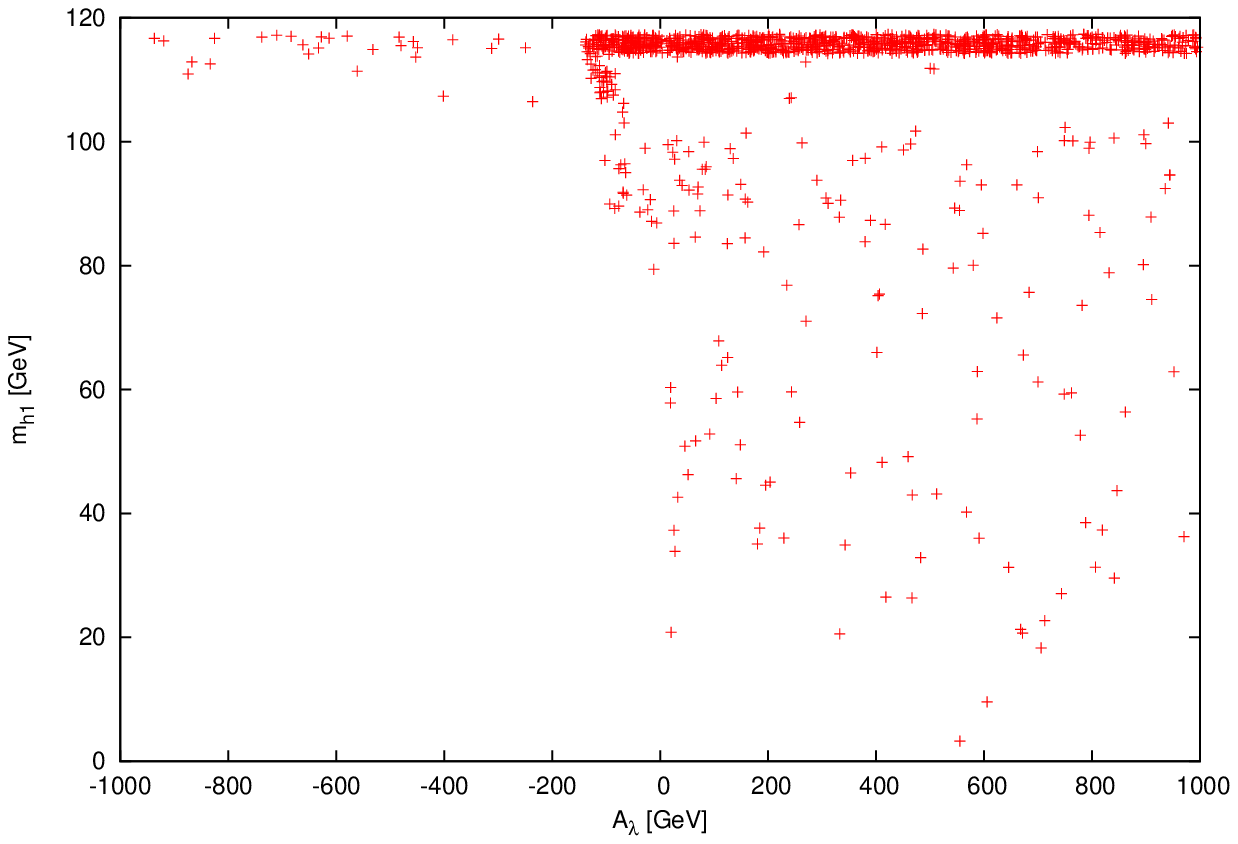}&\includegraphics[scale=0.60]{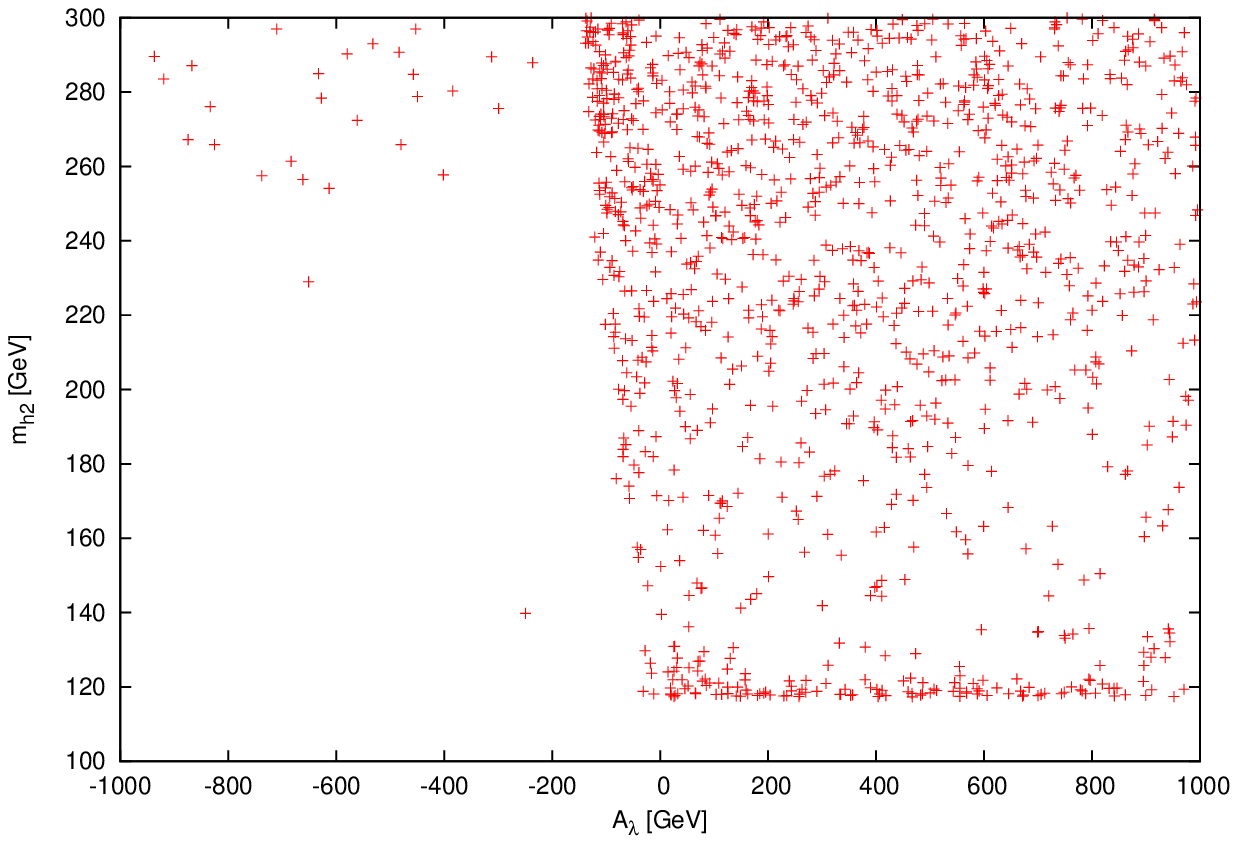}\\
   \includegraphics[scale=0.60]{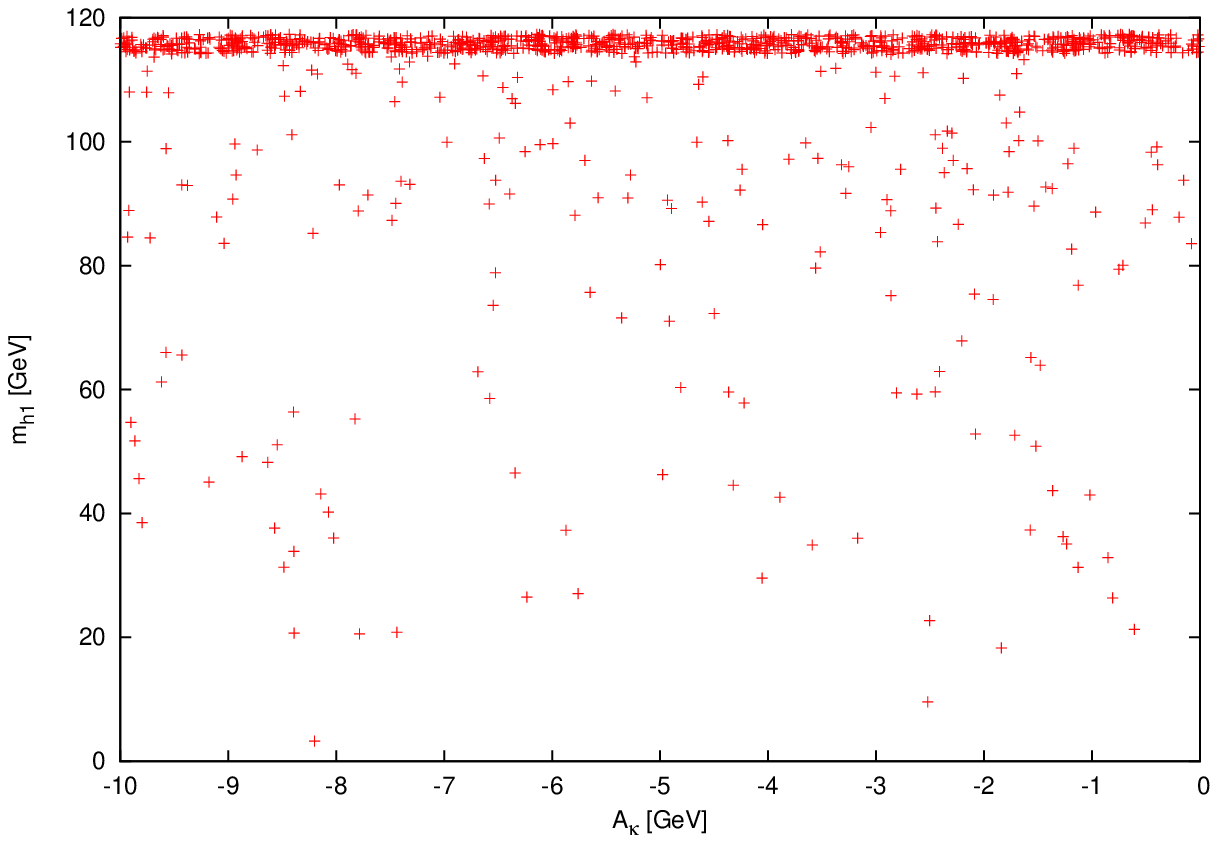}&\includegraphics[scale=0.60]{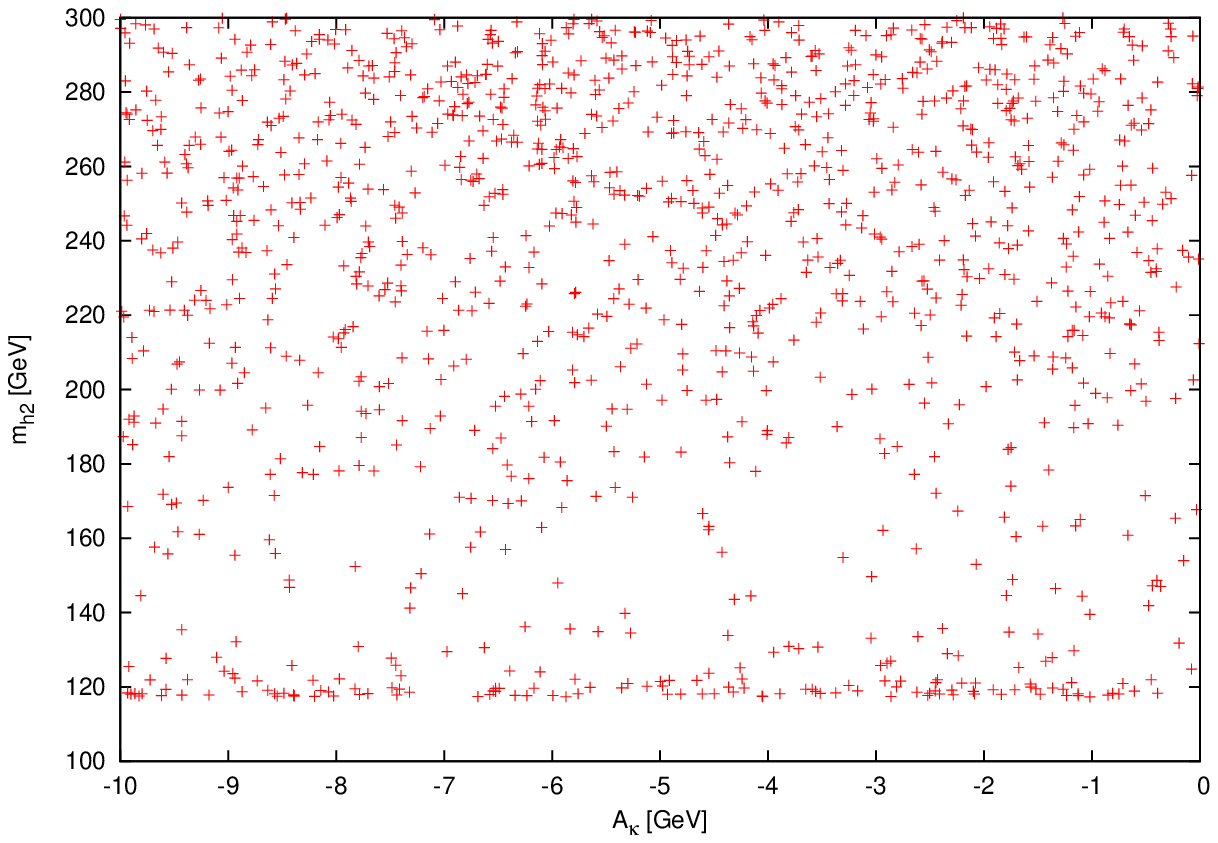}
  \end{tabular}
\label{fig:m1,2-scan1}
\caption{The lightest two scalar Higgs masses $m_{h_1}$ and $m_{h_2}$ as functions of 
 $\mu_{\rm eff}$, $A_\lambda$ and $A_\kappa$. }

\end{figure}

\begin{figure}
 \centering\begin{tabular}{cc}

 \includegraphics[scale=0.60]{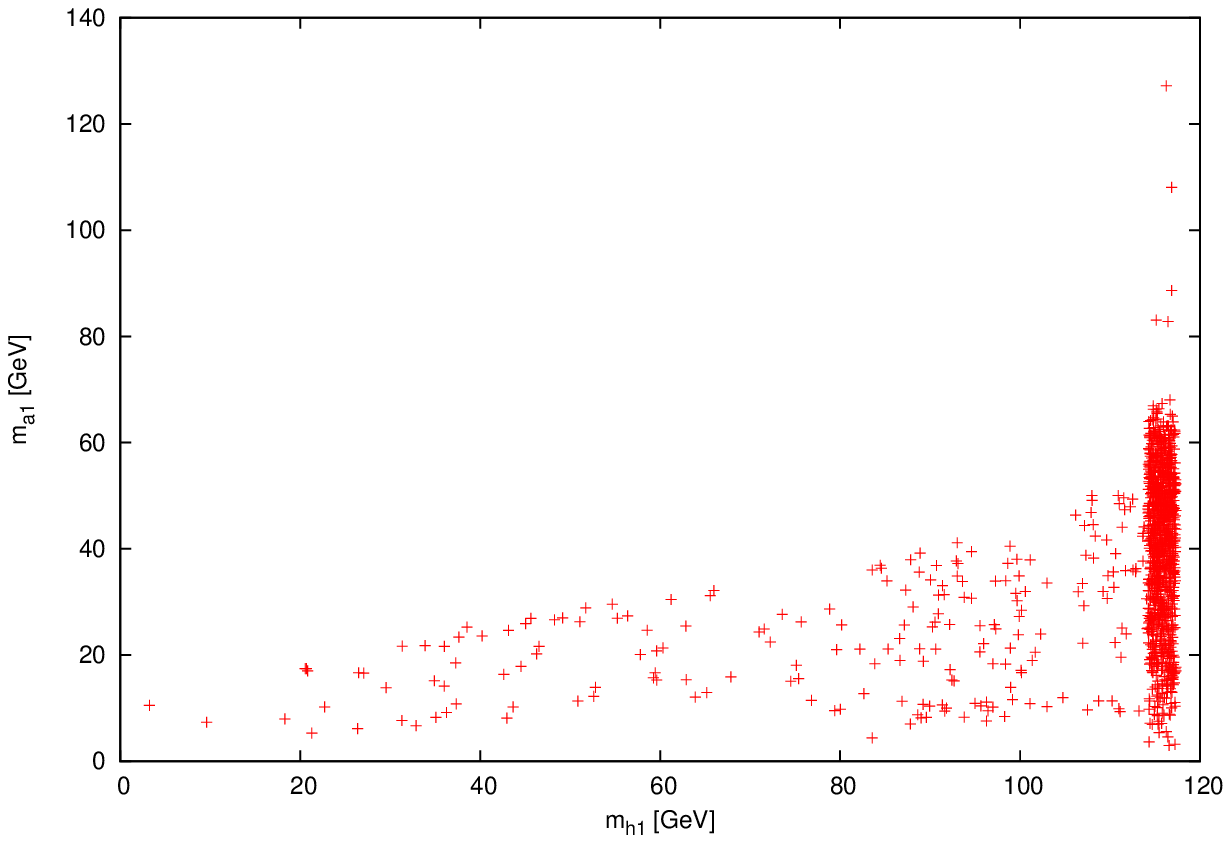}\\
\includegraphics[scale=0.60]{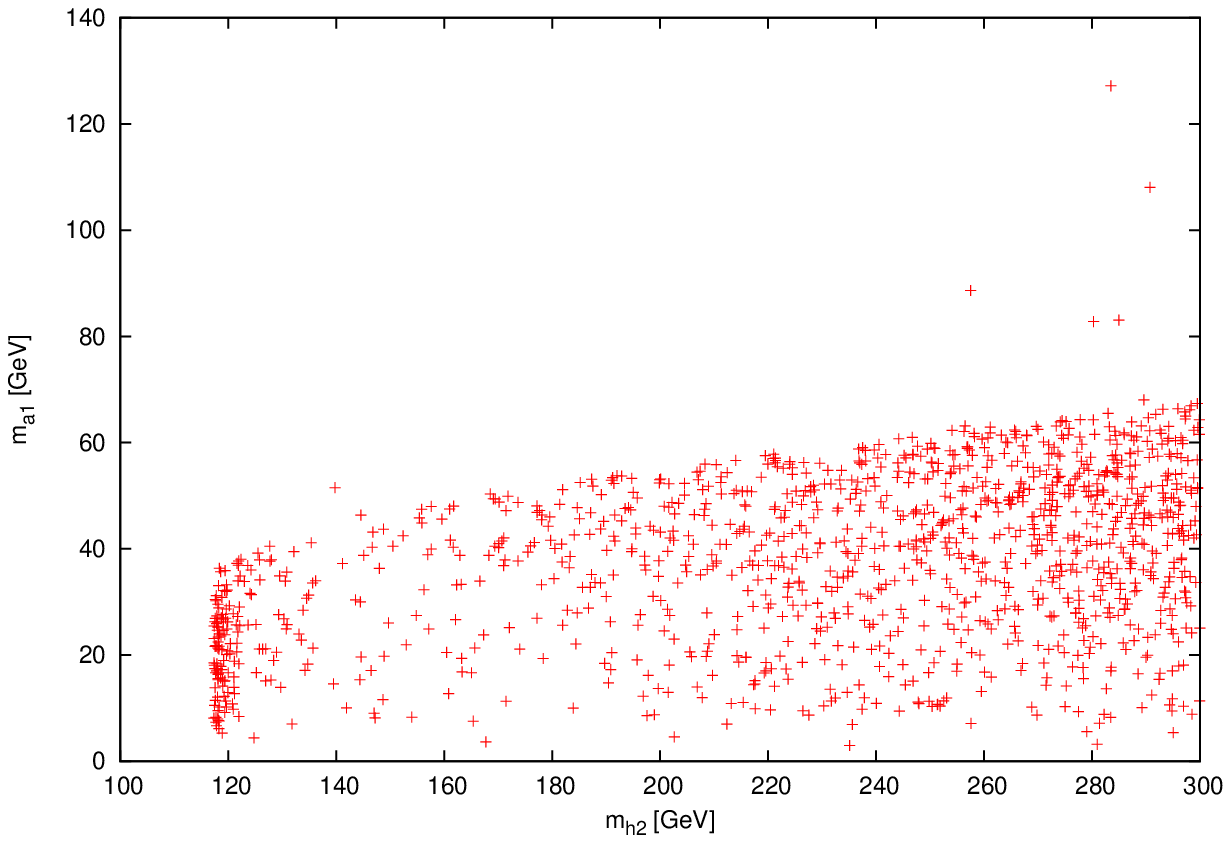}\\
  \includegraphics[scale=0.60]{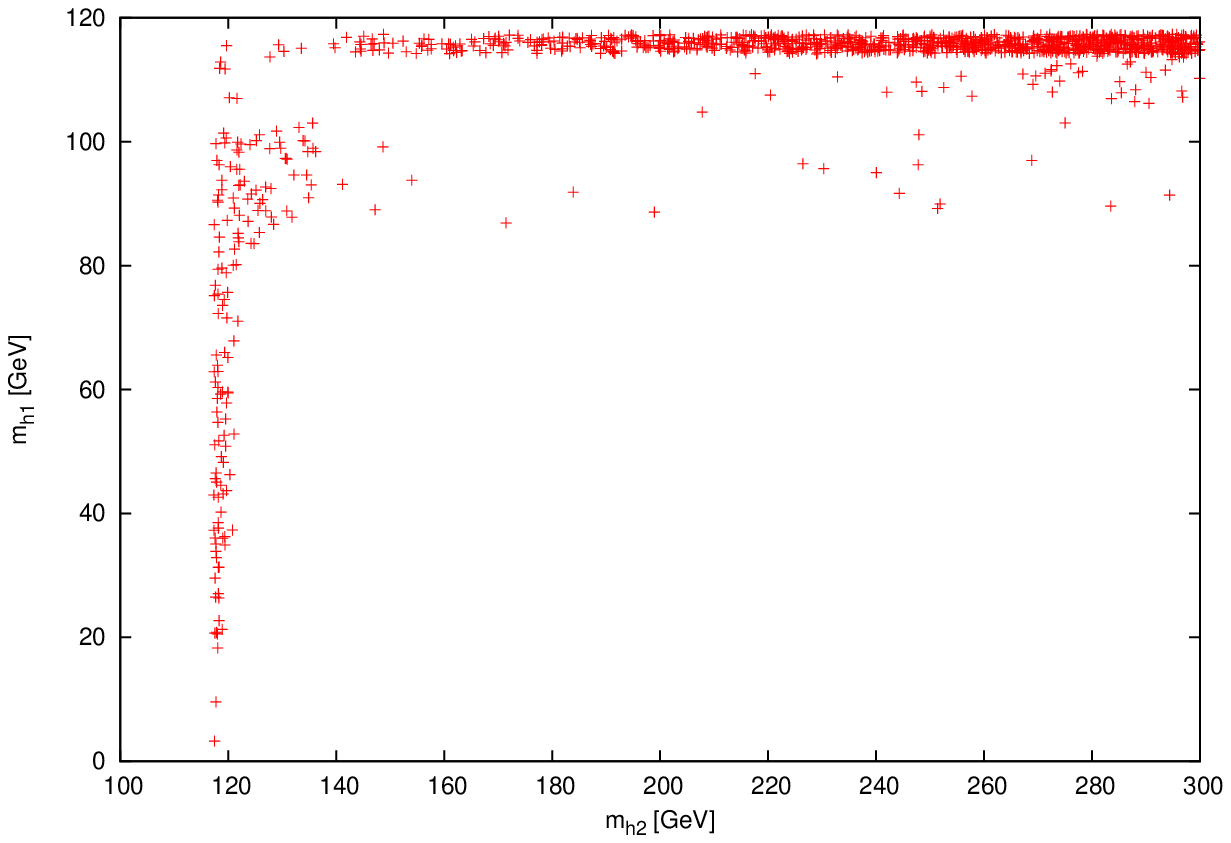}
 \end{tabular}
\label{fig:m1,2-scan2}
\caption{The correlations between the lightest CP-odd Higgs mass, $m_{a_1}$
and the lightest two CP-even Higgs masses, $m_{h_1}$ and $m_{h_2}$ and between the latter two.}

\end{figure}

\begin{figure}
 \centering\begin{tabular}{cc}
  \includegraphics[scale=0.4]{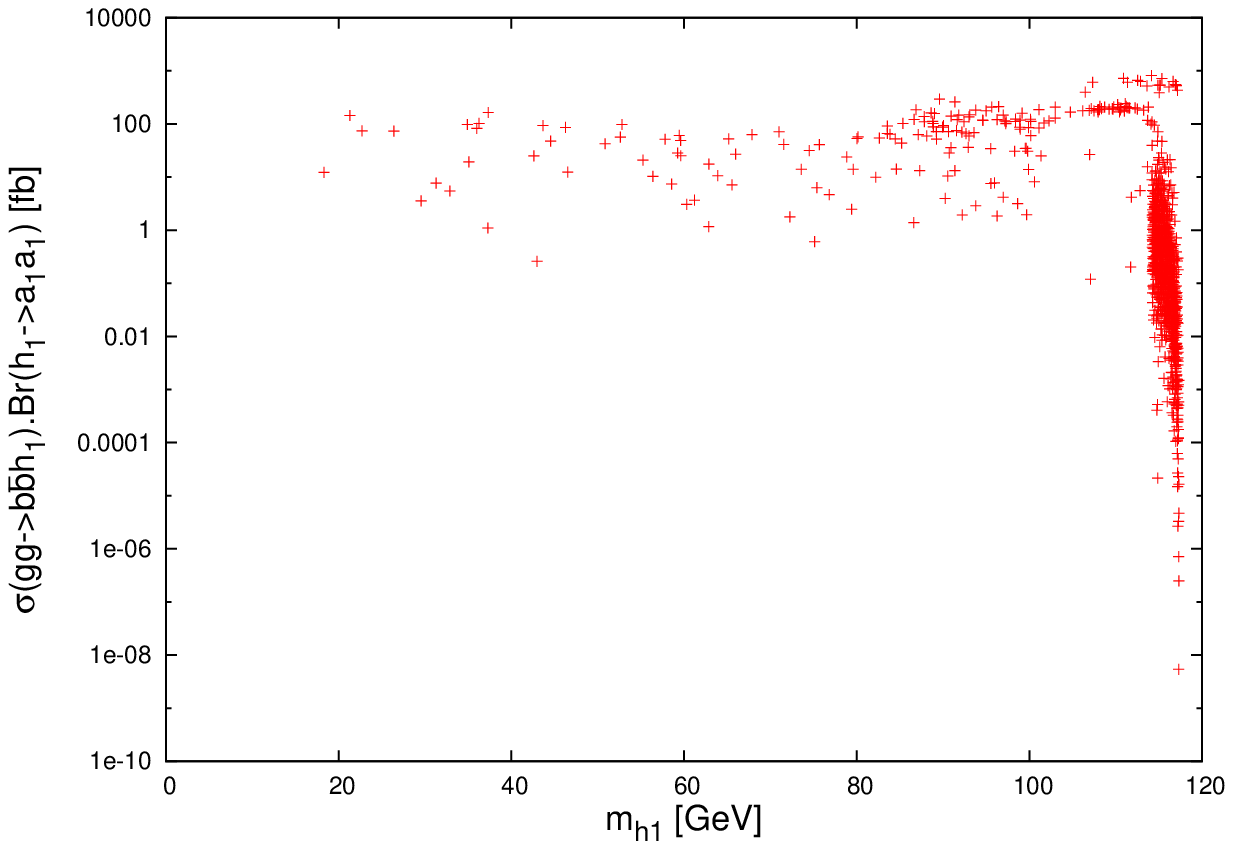}\includegraphics[scale=0.4]{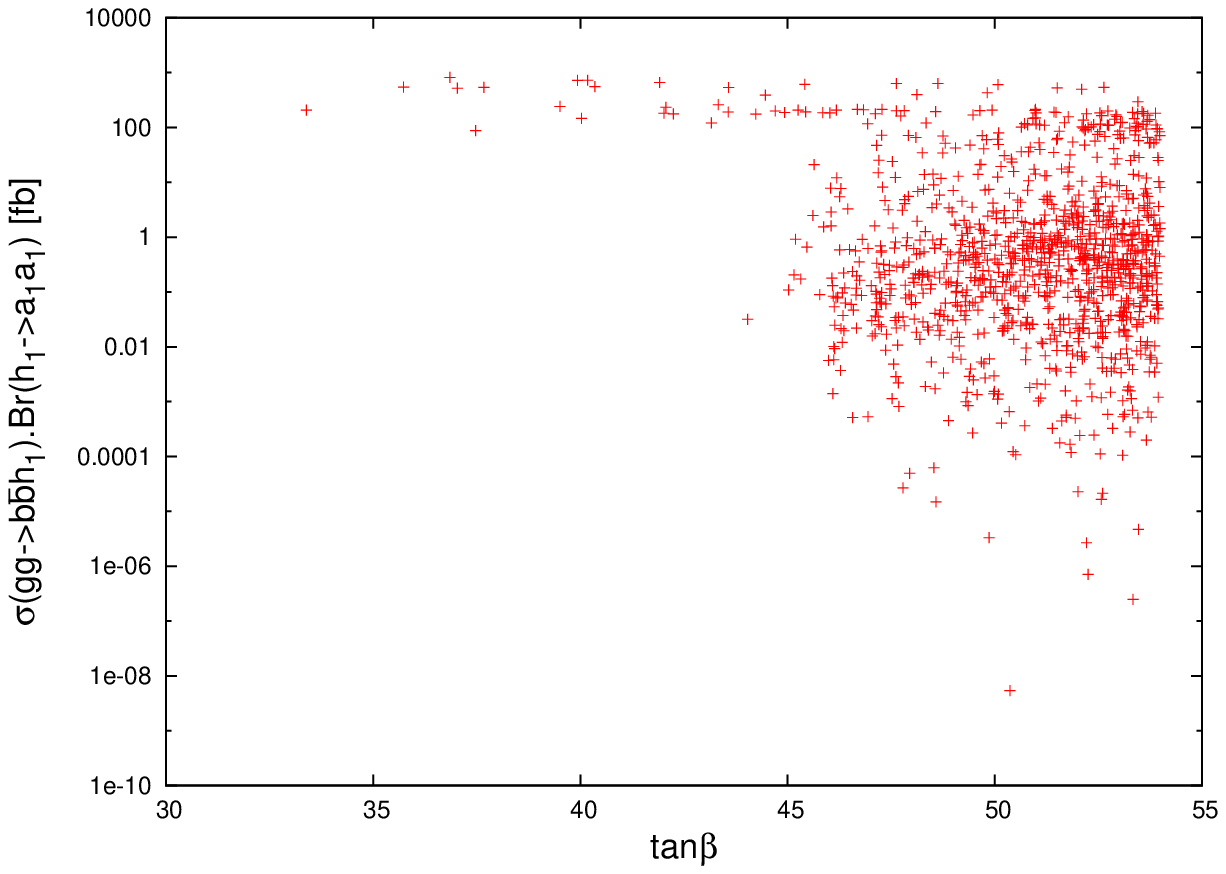}
\includegraphics[scale=0.4]{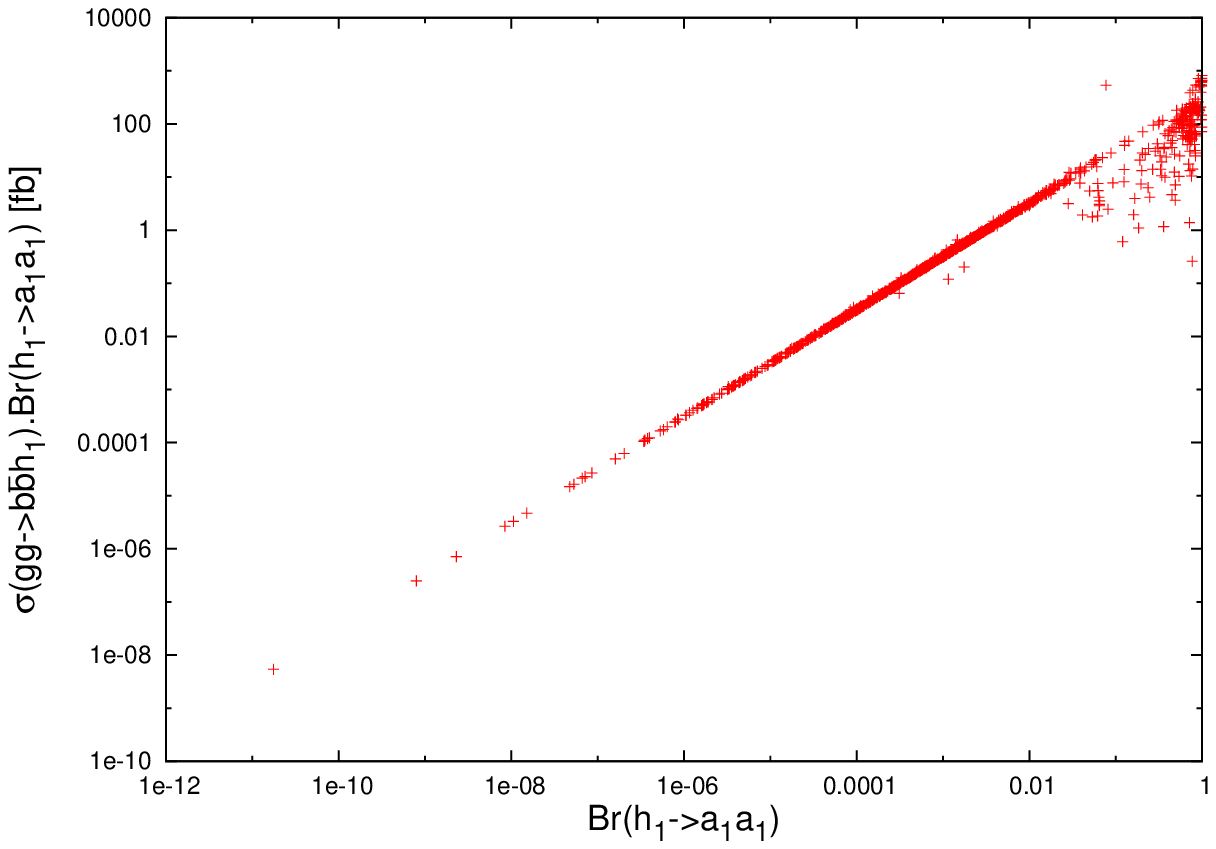}\\
  \includegraphics[scale=0.4]{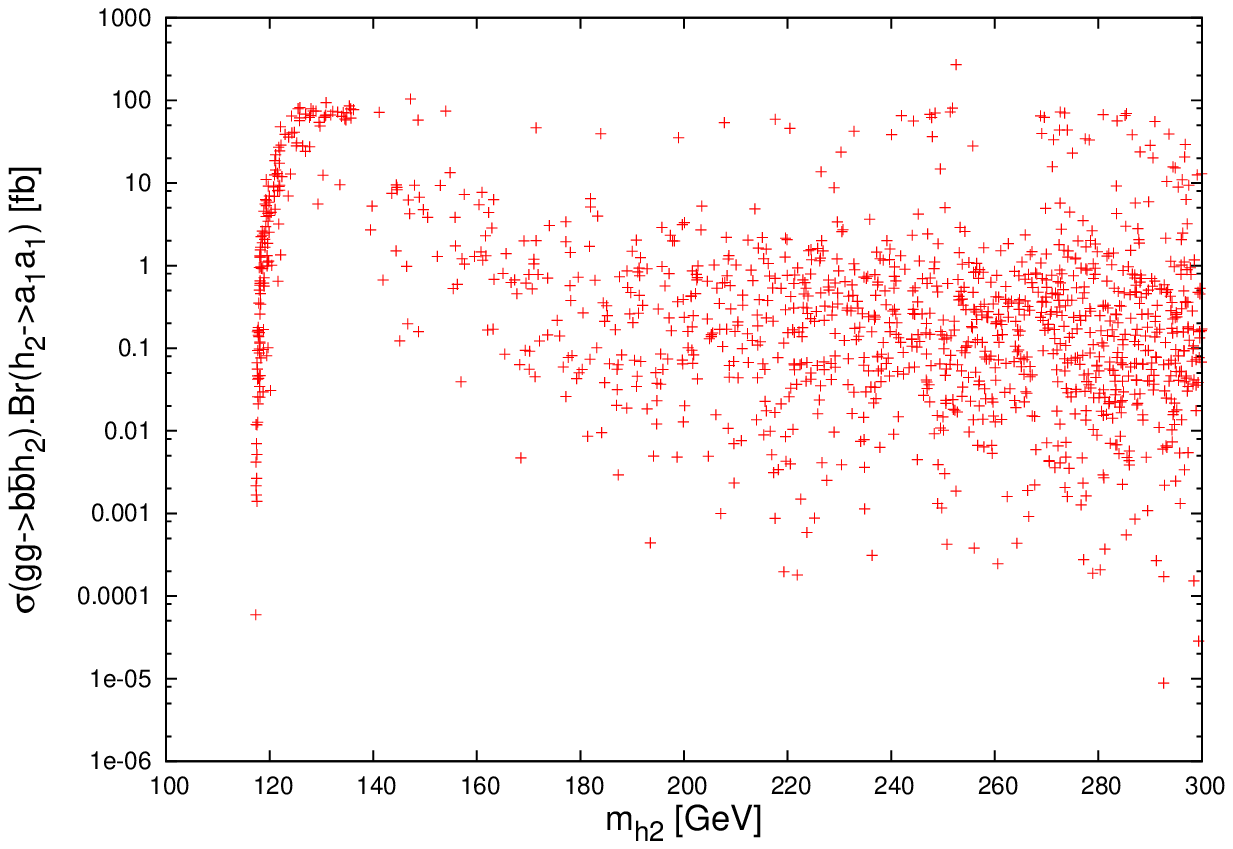}\includegraphics[scale=0.4]{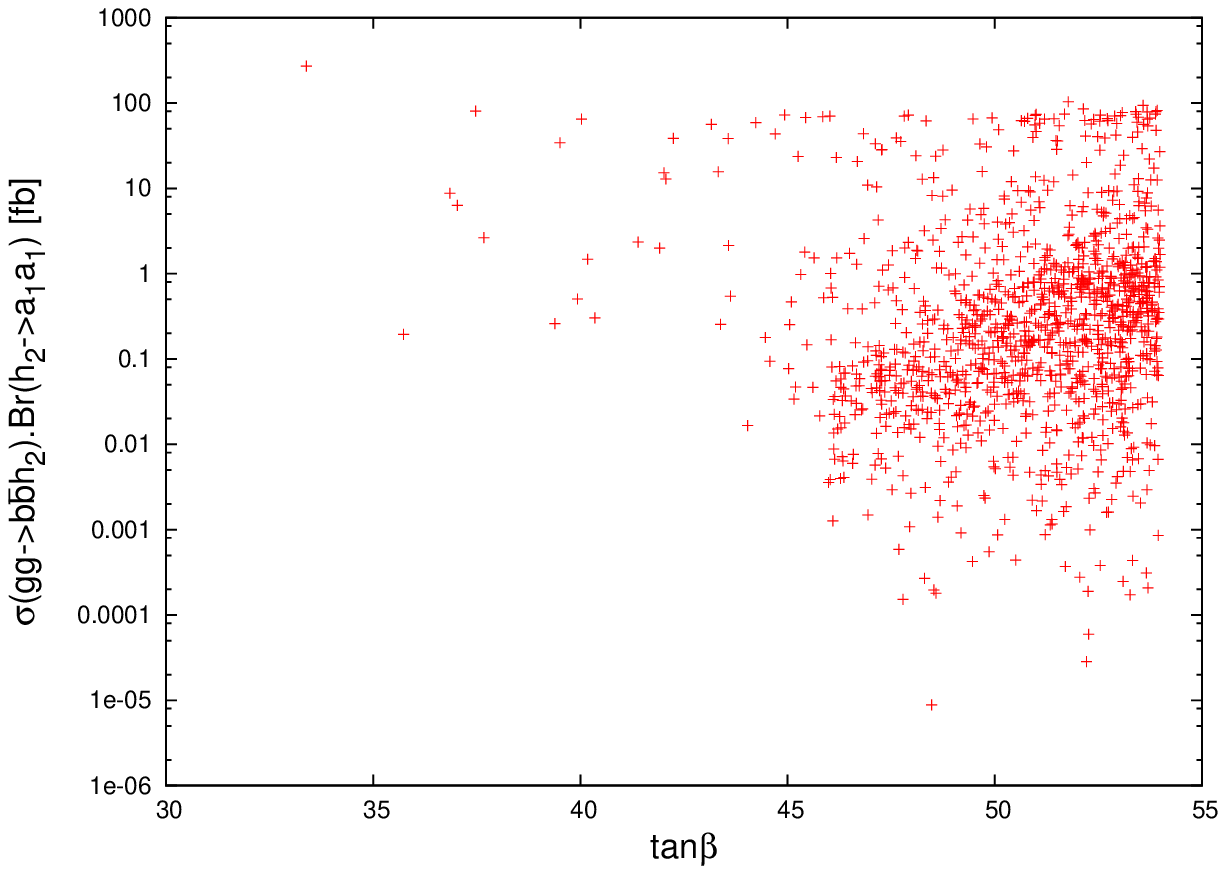}
\includegraphics[scale=0.4]{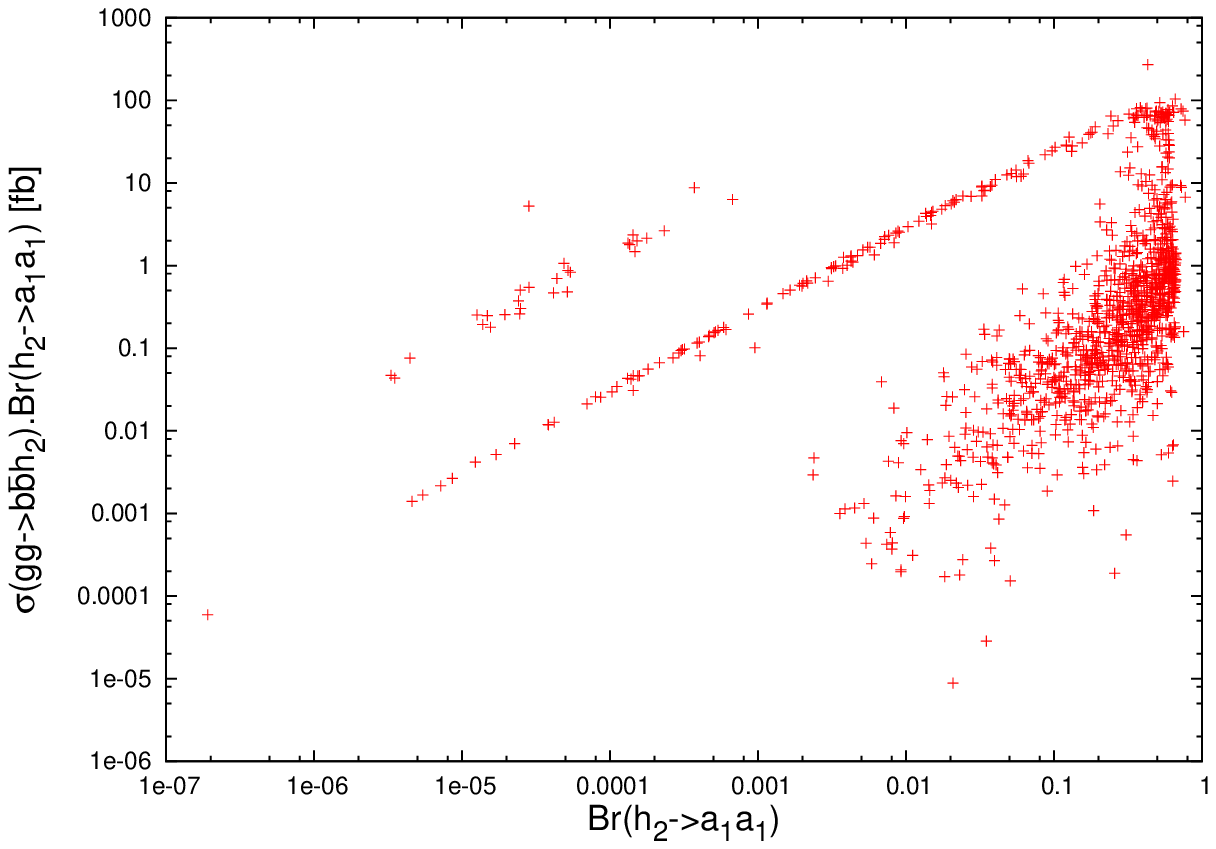}\\
  \includegraphics[scale=0.4]{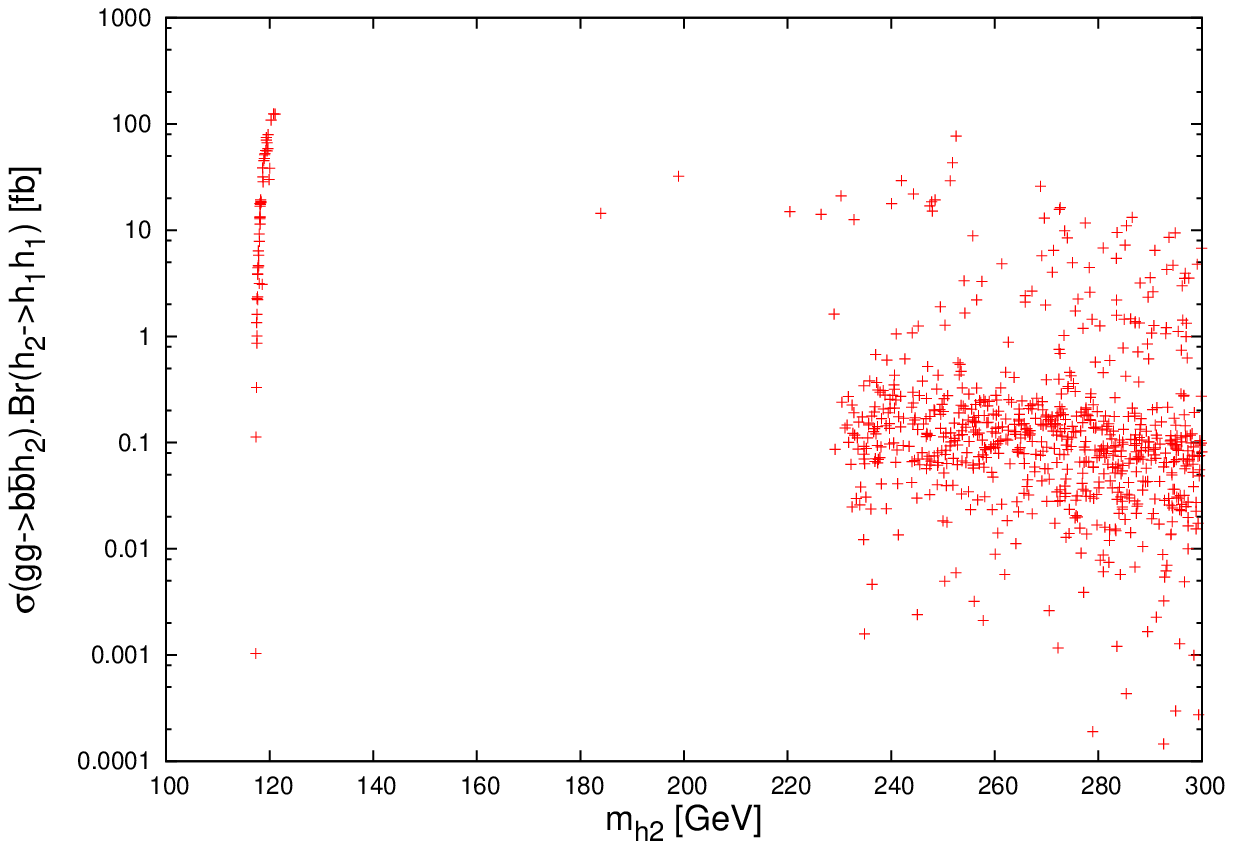}\includegraphics[scale=0.4]{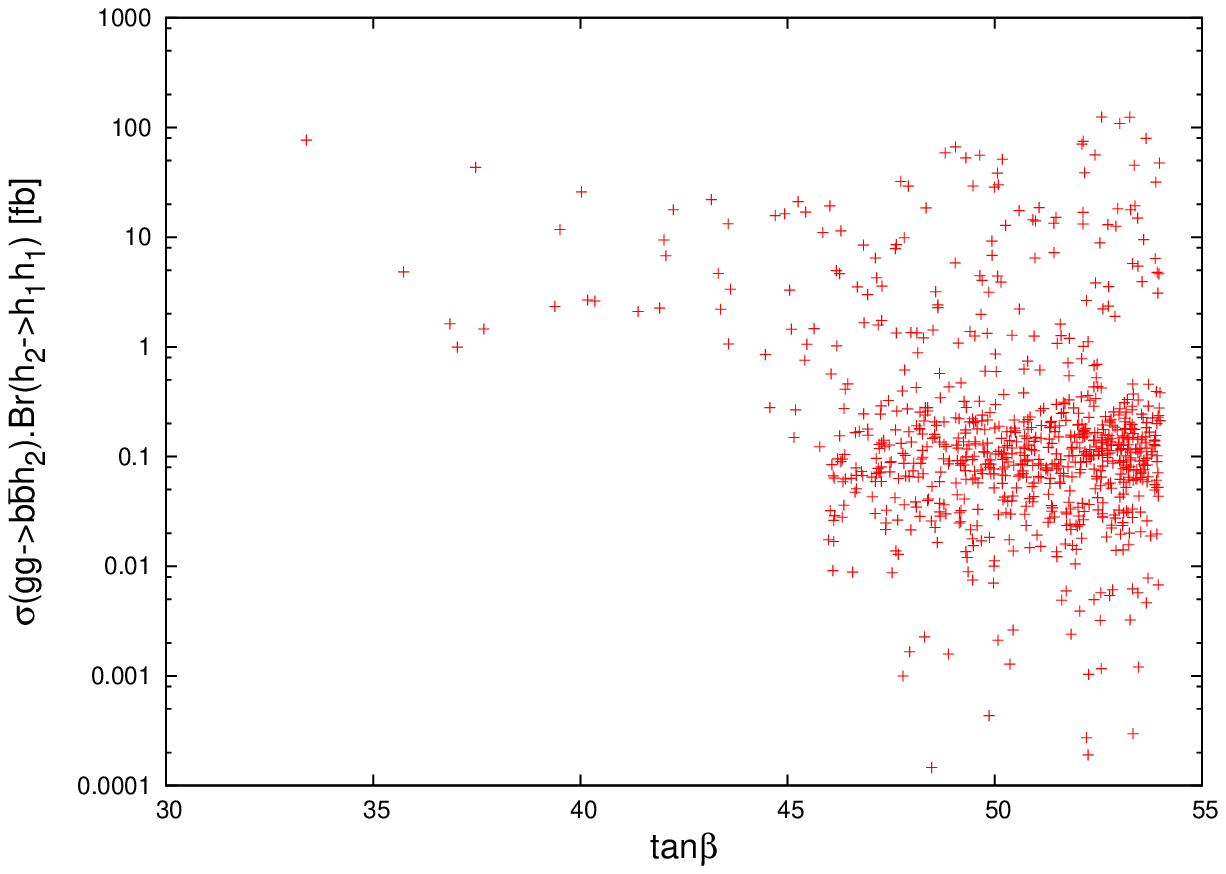}
\includegraphics[scale=0.4]{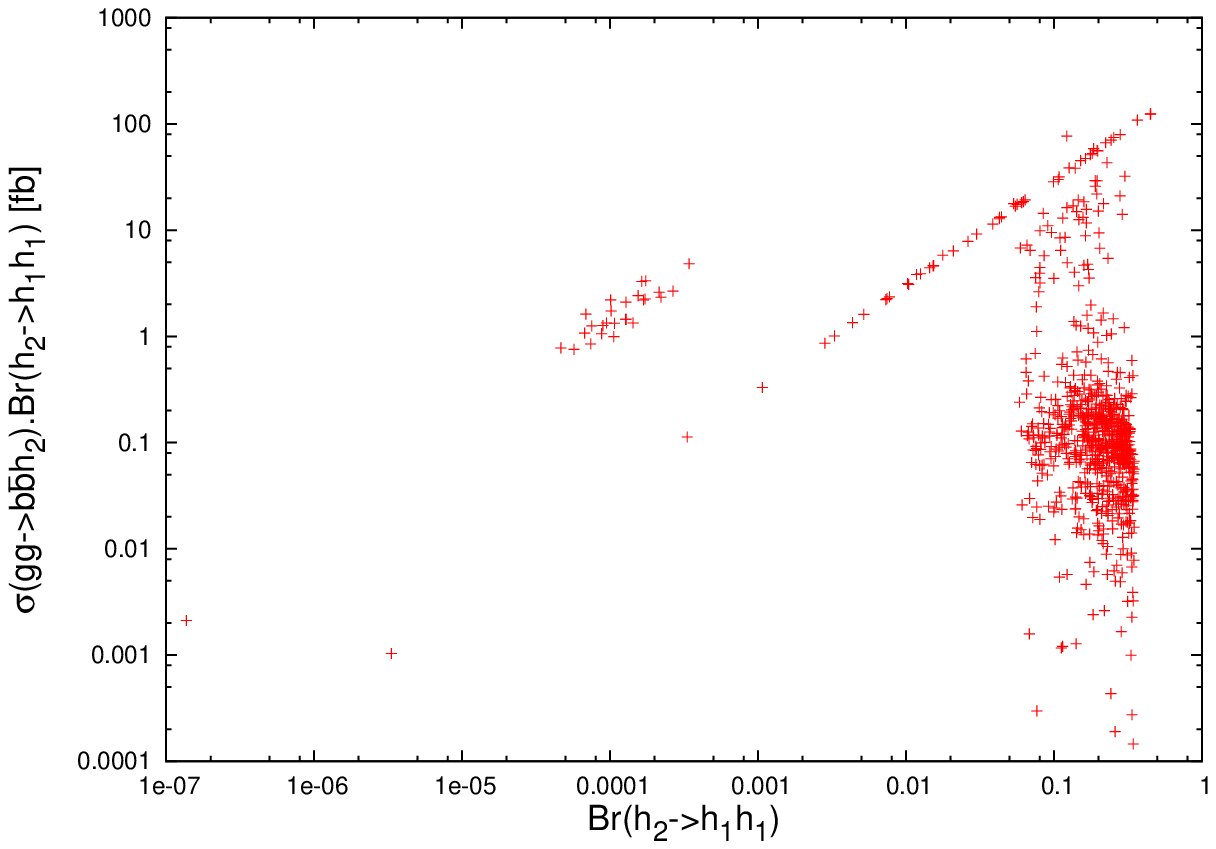}

 \end{tabular}
\label{fig:sigma-scan3}
\caption{The rates for $\sigma(pp\to b\bar b {h_1})~{\rm BR}(h_1\to a_1a_1)$, 
$\sigma(pp\to b\bar b {h_2})~{\rm BR}(h_2\to a_1a_1)$ 
 and $\sigma(pp\to b\bar b {h_2})~{\rm BR}(h_2\to h_1h_1)$ as functions 
of corresponding Higgs masses, of tan$\beta$ and of corresponding BRs.}

\end{figure}

\begin{figure}
 \centering\begin{tabular}{cc}
  \includegraphics[scale=0.60]{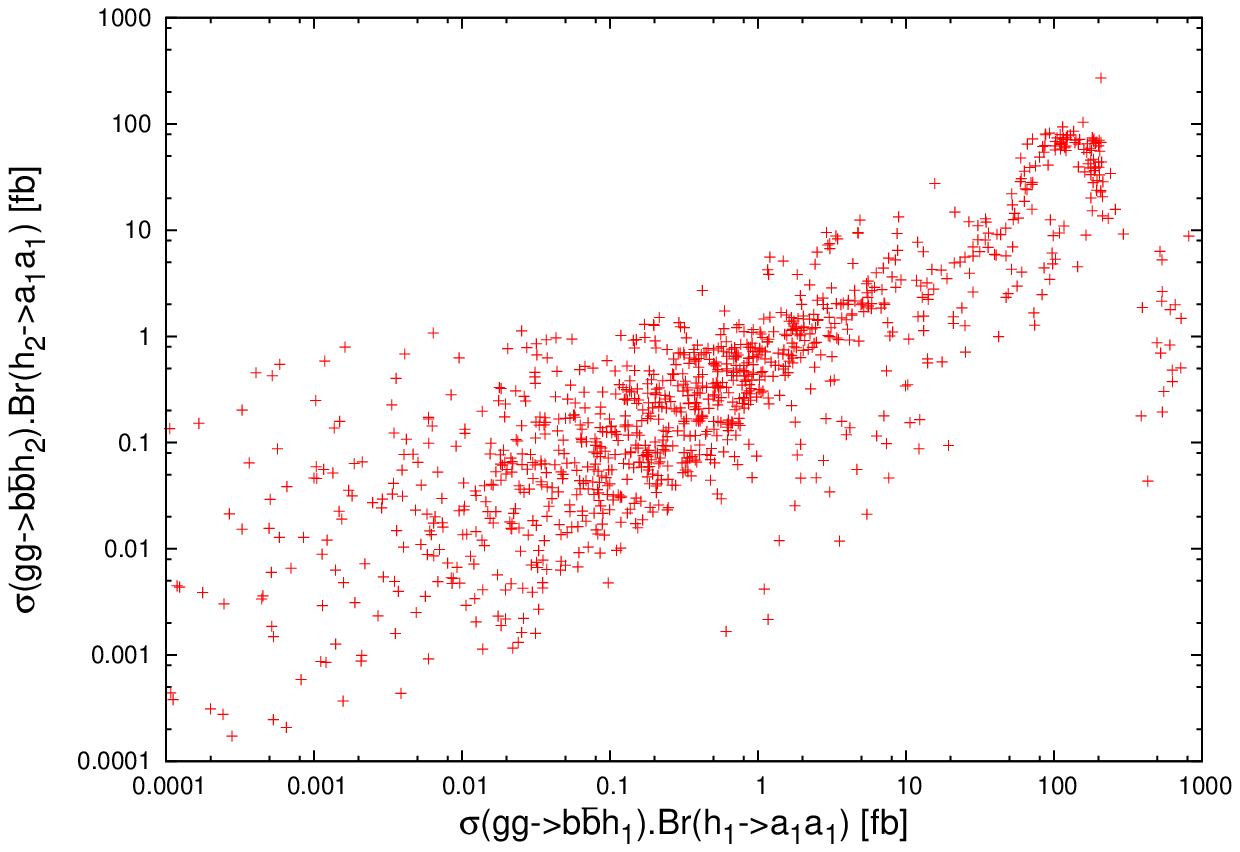}\\
   \includegraphics[scale=0.60]{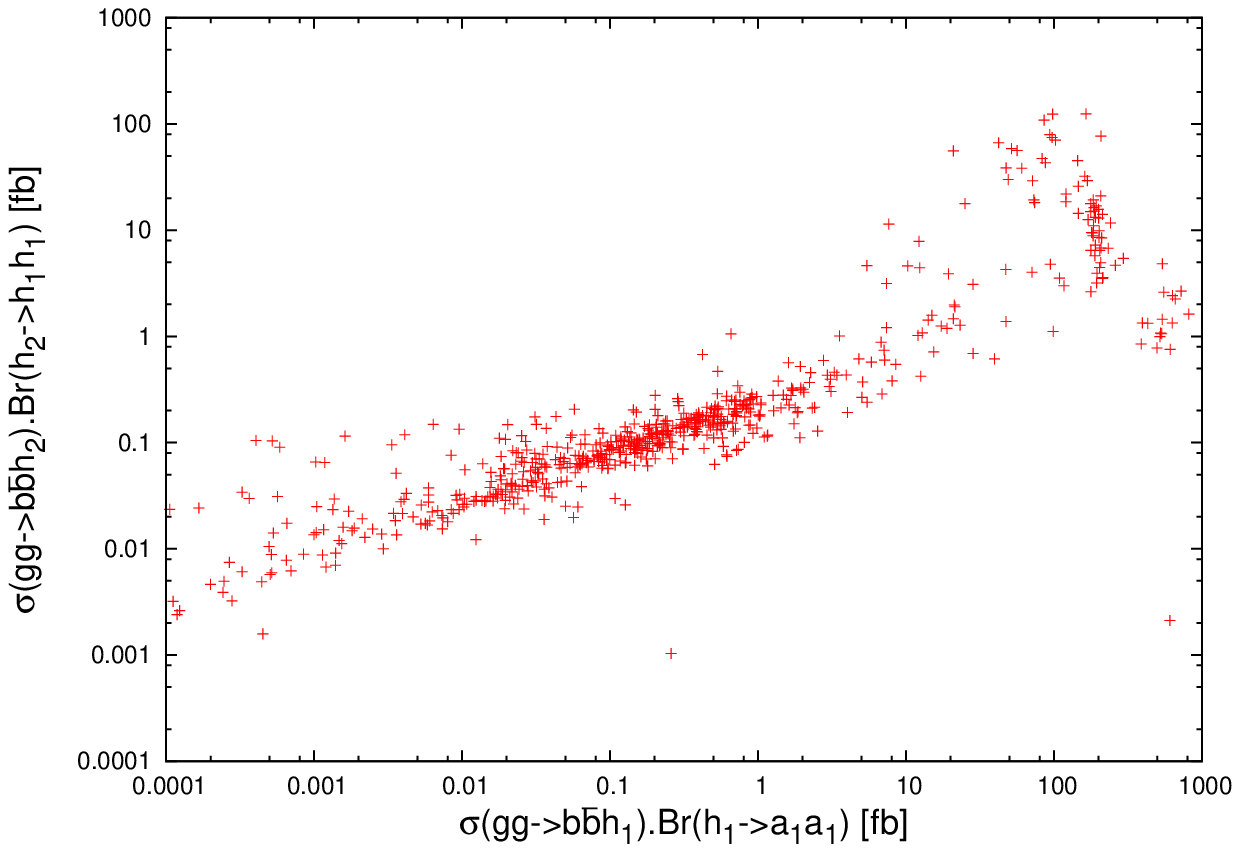}\\
  \includegraphics[scale=0.60]{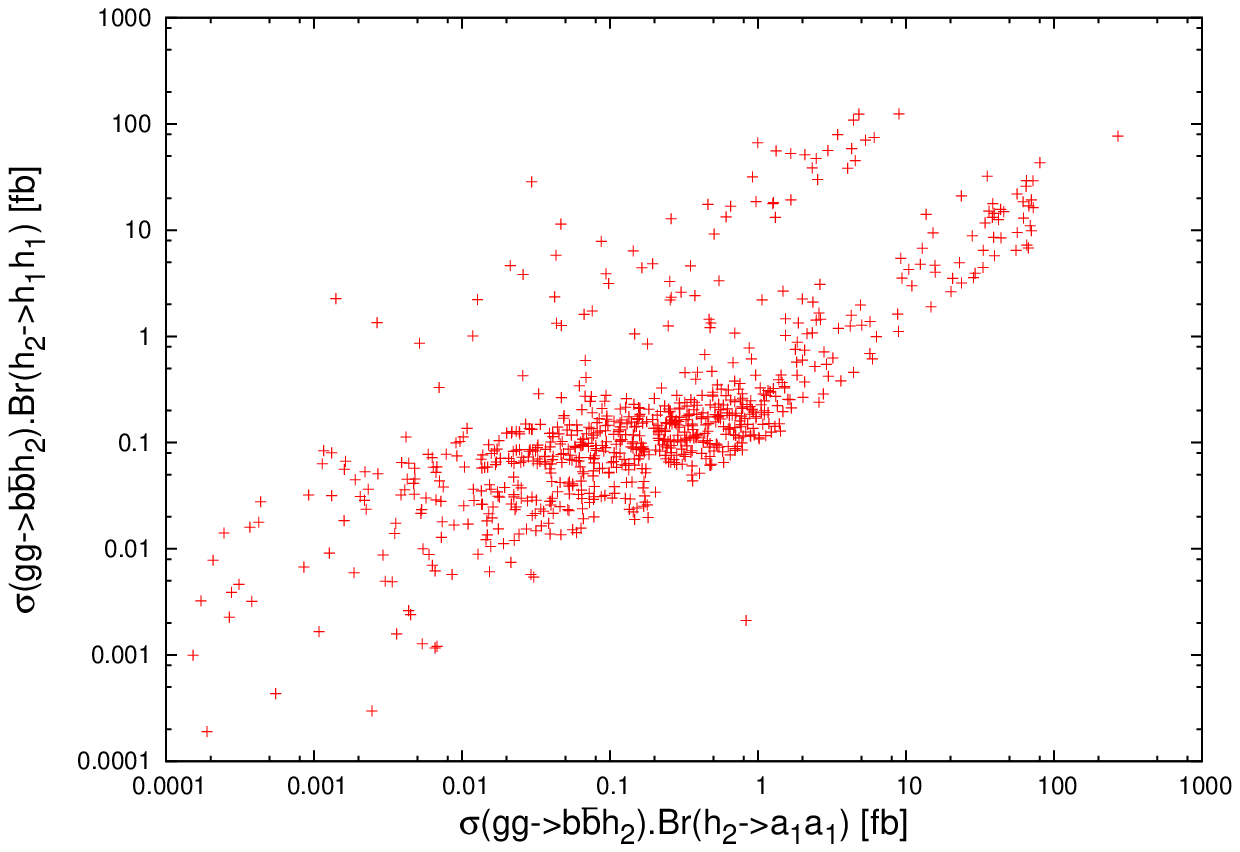}
     
 \end{tabular}
   \label{fig:sigma-correlations}
\caption{The rates for 
$\sigma(pp\to b\bar b {h_1})~{\rm BR}(h_1\to a_1a_1)$  versus  
$\sigma(pp\to b\bar b {h_2})~{\rm BR}(h_2\to a_1a_1)$, for
$\sigma(pp\to b\bar b {h_1})~{\rm BR}(h_1\to a_1a_1)$  versus  
$\sigma(pp\to b\bar b {h_2})~{\rm BR}(h_2\to h_1h_1)$  and for
$\sigma(pp\to b\bar b {h_1})~{\rm BR}(h_2\to a_1a_1)$  versus  
$\sigma(pp\to b\bar b {h_2})~{\rm BR}(h_2\to h_1h_1)$.}

\end{figure}

\begin{figure}
 \centering\begin{tabular}{cc}
  \includegraphics[scale=0.60]{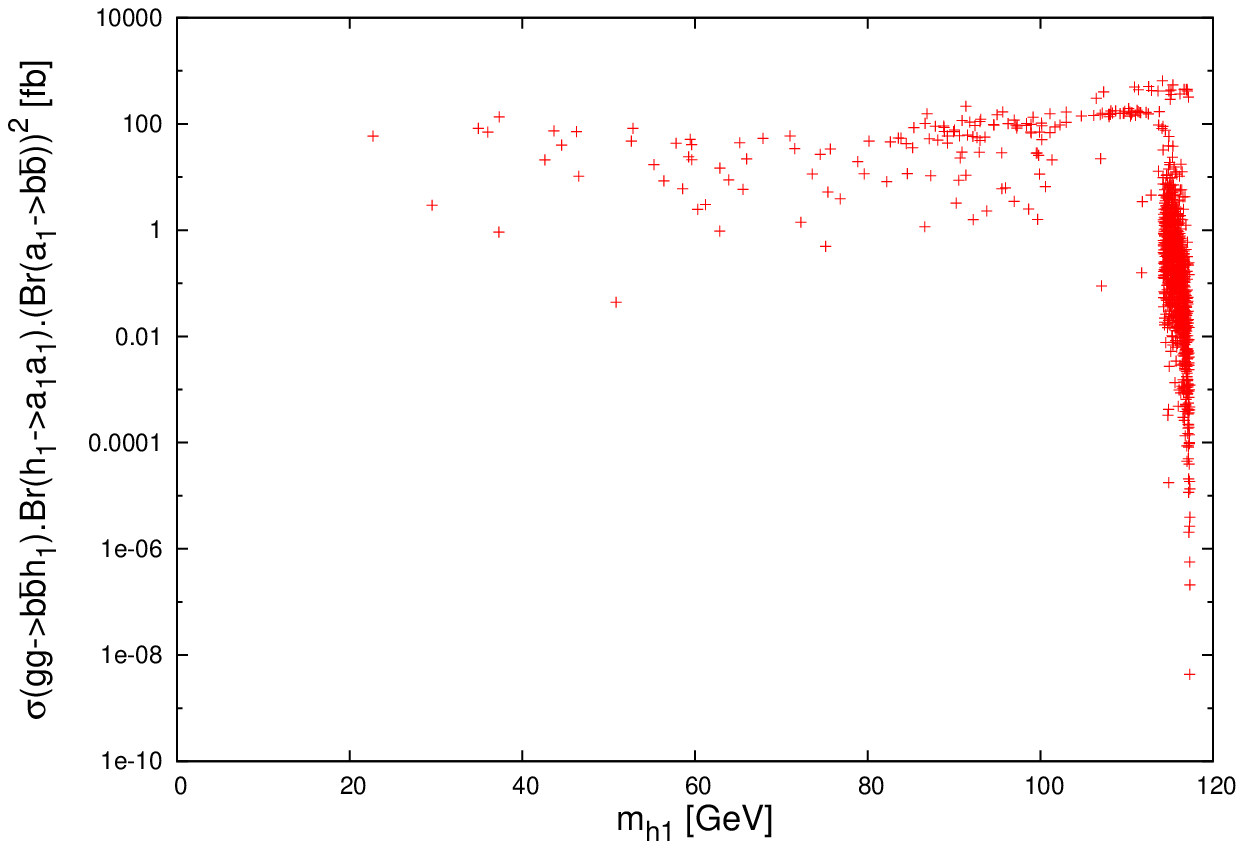}&\includegraphics[scale=0.60]{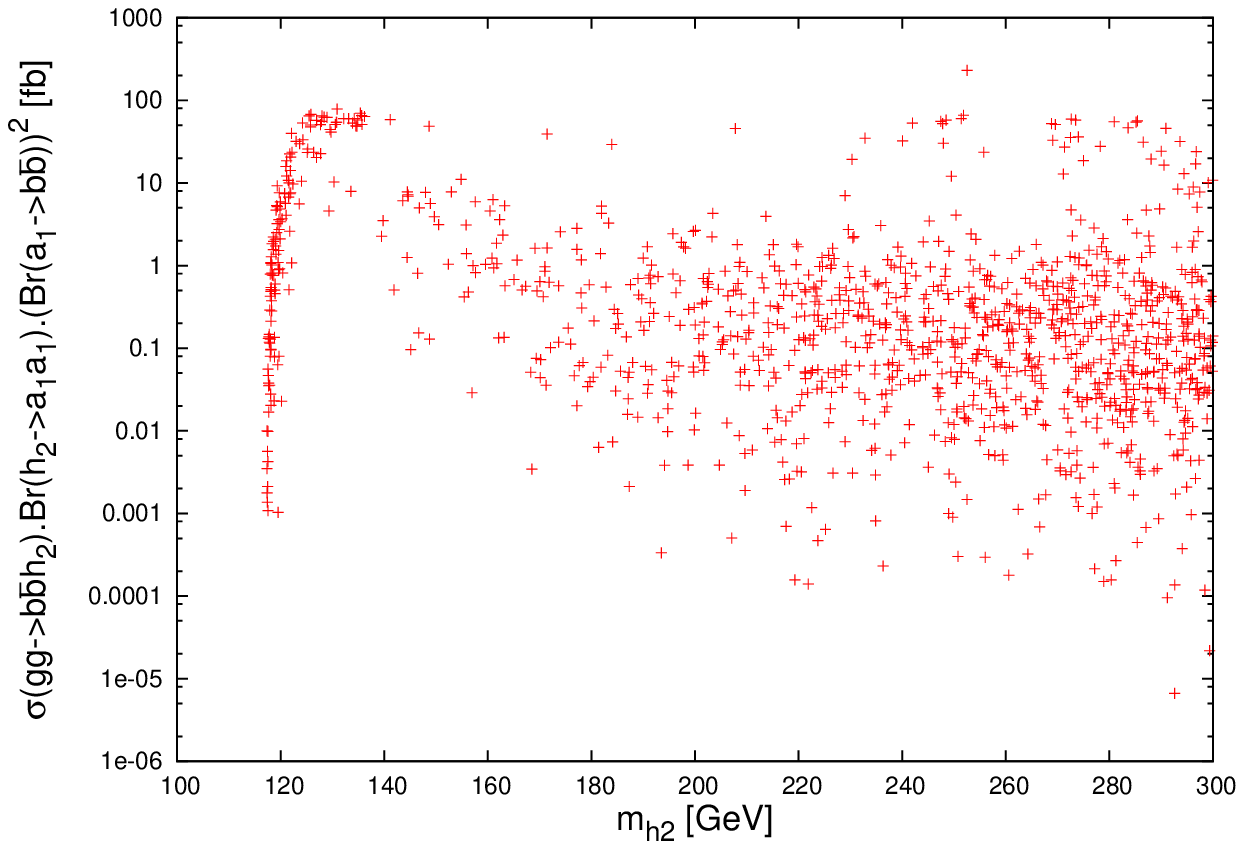}\\
   \includegraphics[scale=0.60]{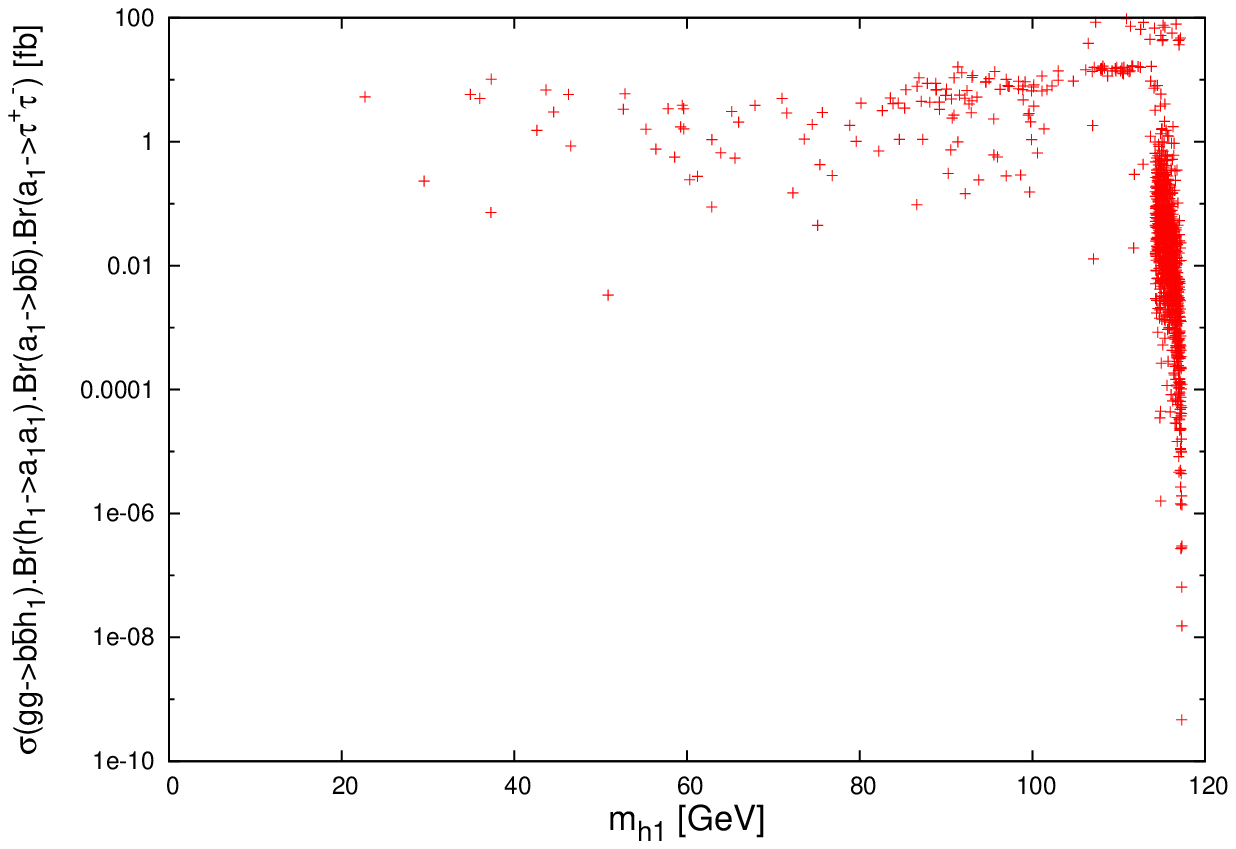}&\includegraphics[scale=0.60]{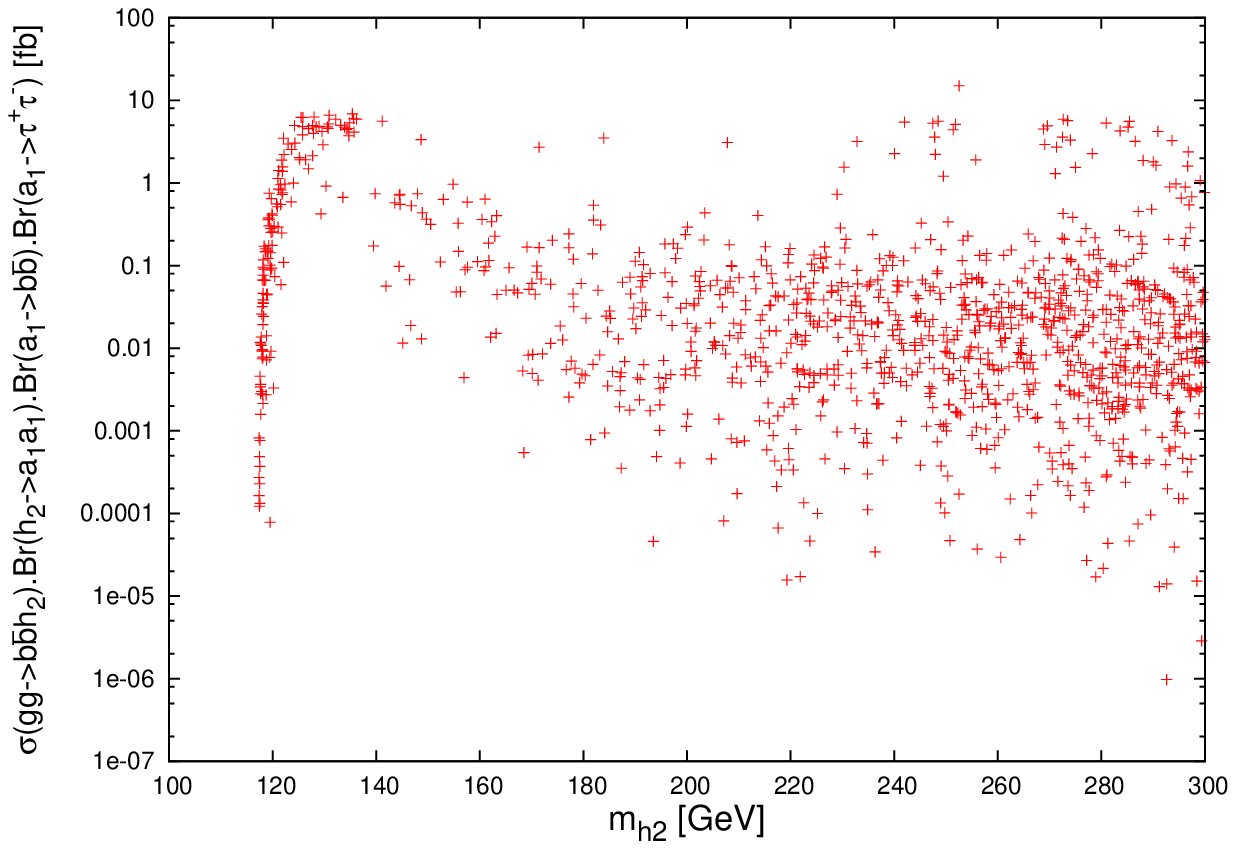}\\
    \includegraphics[scale=0.60]{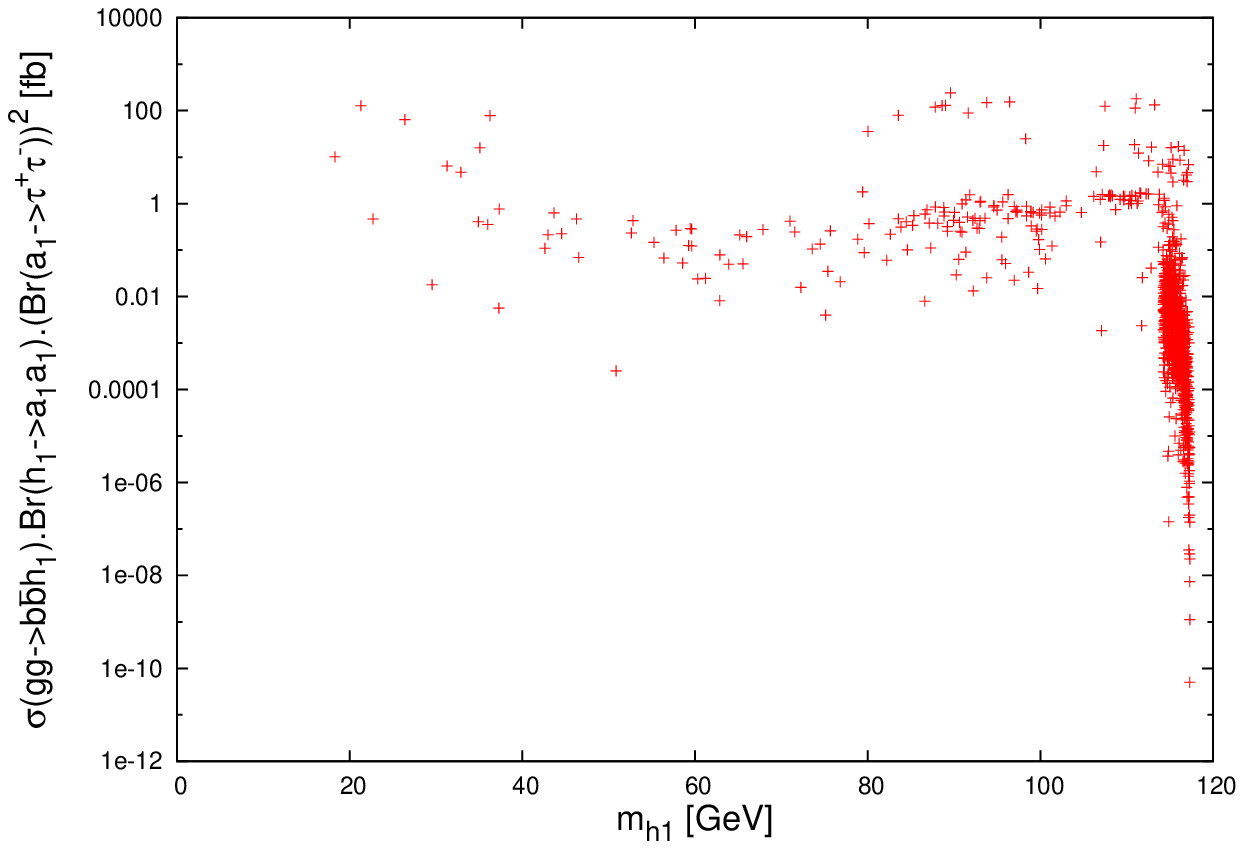}&\includegraphics[scale=0.60]{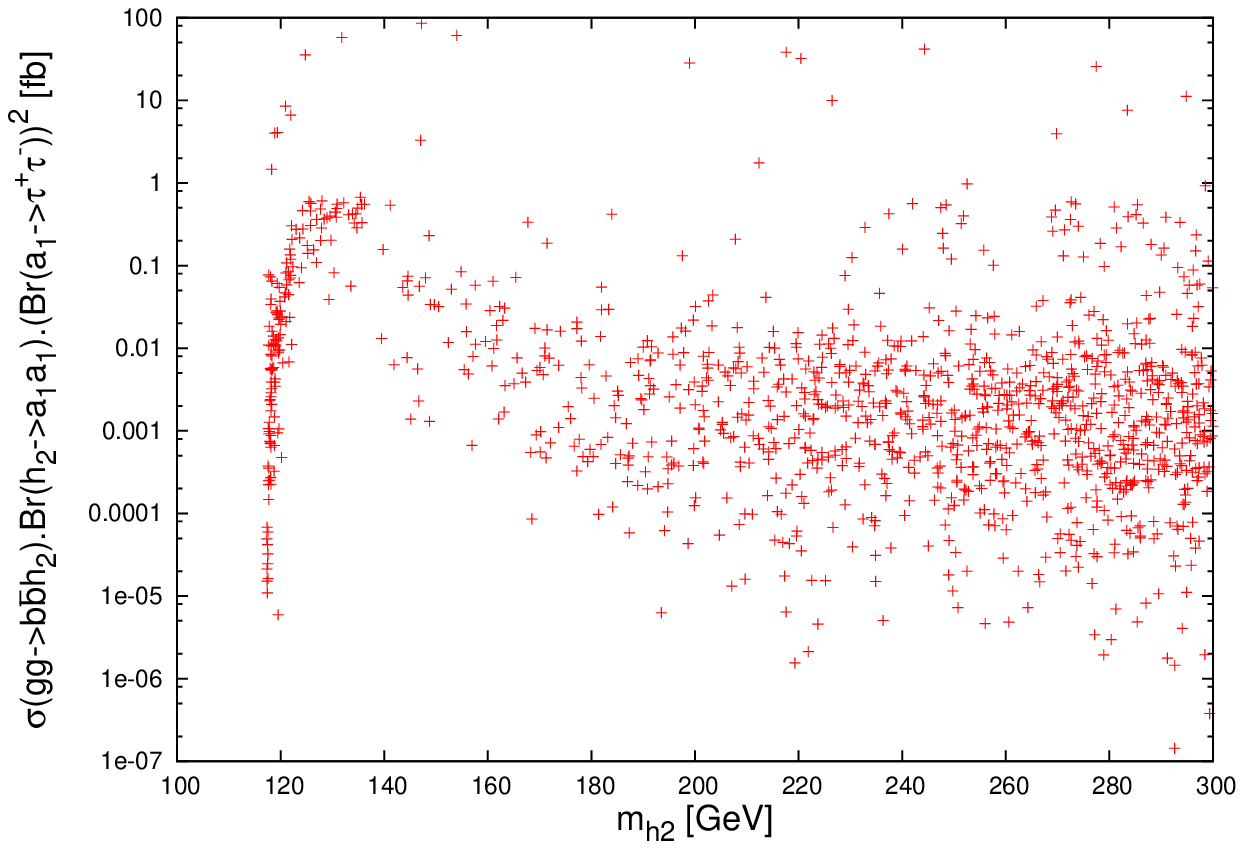}
   \end{tabular}
   \label{fig:sigma-scan4}
\caption{The signal rates for $\sigma(pp\to b\bar b {h_1})~{\rm BR}(h_1\to a_1a_1)$ and 
$\sigma(pp\to b\bar b {h_2})~{\rm BR}(h_2\to a_1a_1)$ times BR($a_1a_1\to b\bar bb\bar b$), BR($a_1a_1\to b\bar b\tau^+\tau^-$)
and BR($a_1a_1\to \tau^+\tau^-\tau^+\tau^-$)   
 as functions of $m_{h_1}$ and $m_{h_2}$.}

\end{figure}

\begin{figure}
 \centering\begin{tabular}{cc}
  \includegraphics[scale=0.60]{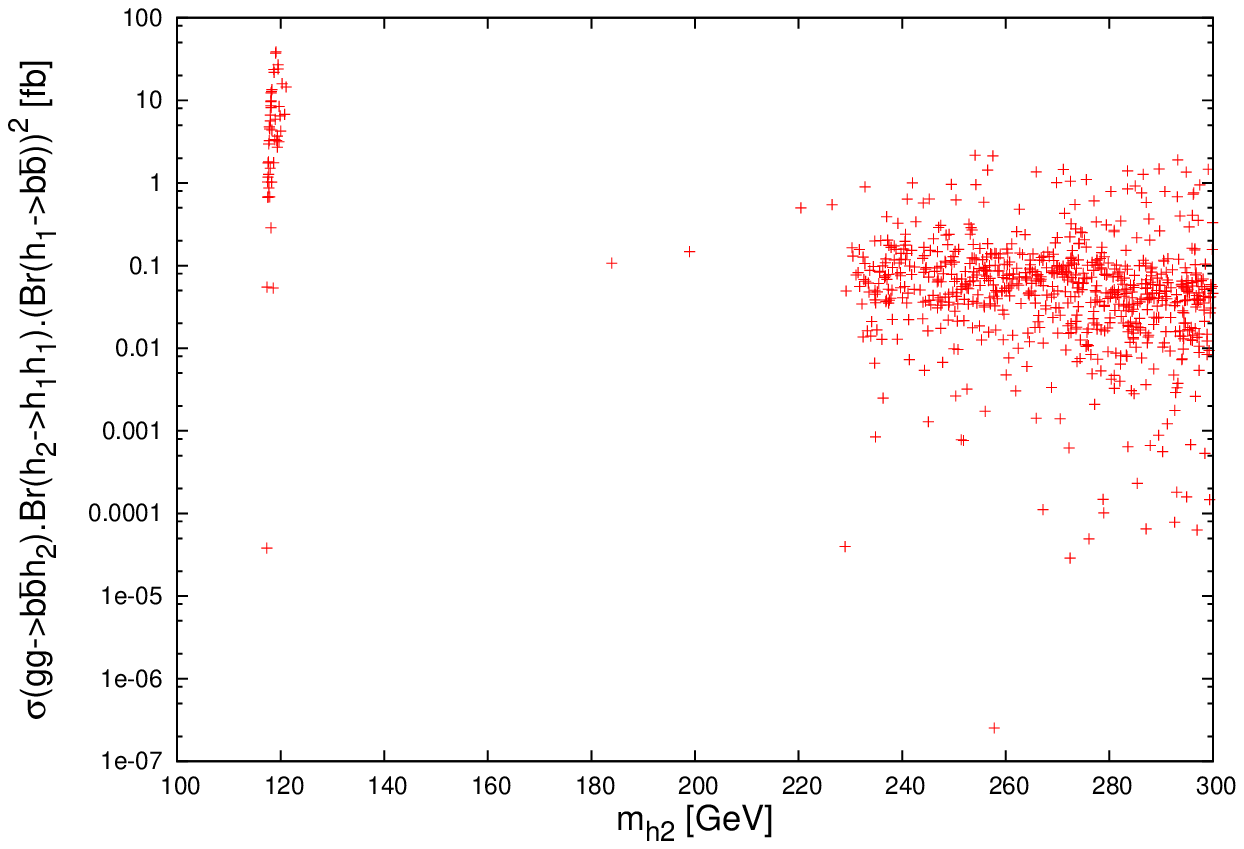}\\
   \includegraphics[scale=0.60]{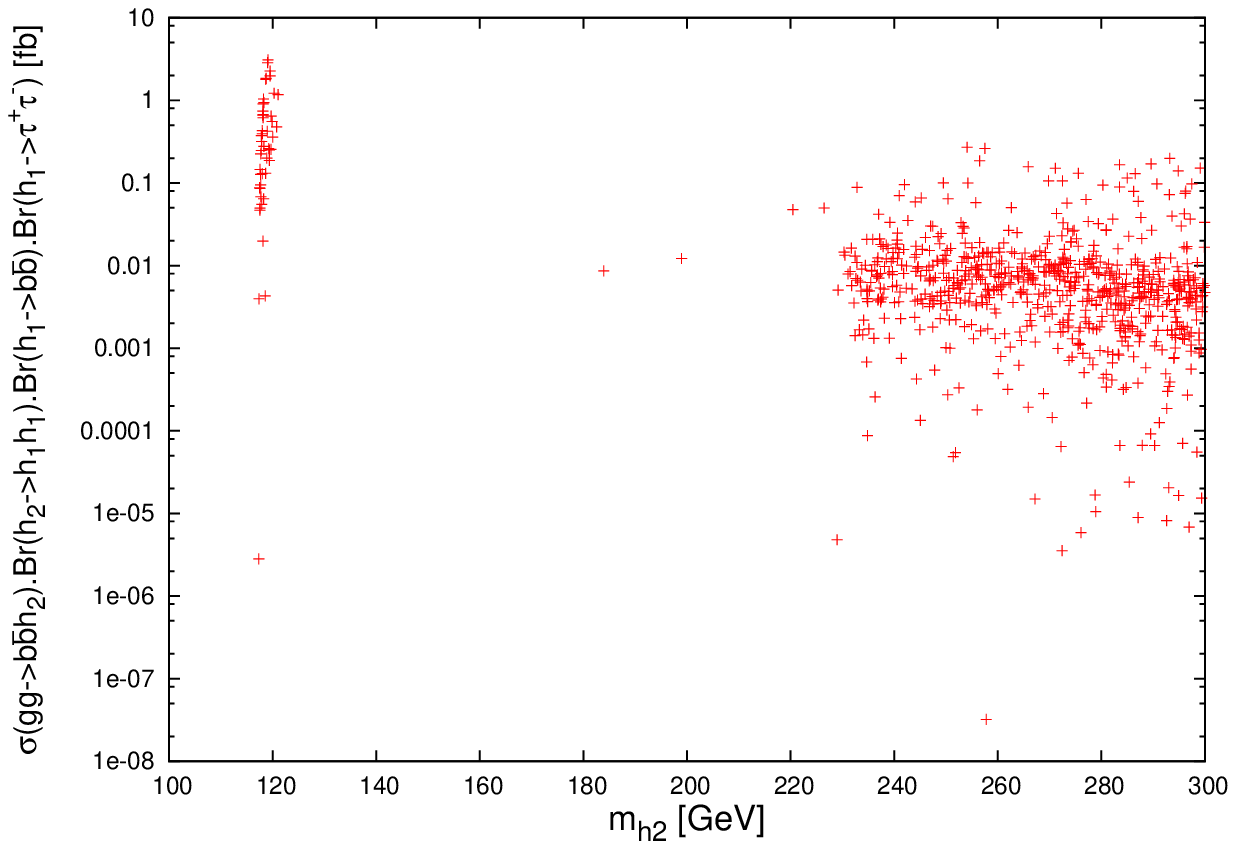}\\
  \includegraphics[scale=0.60]{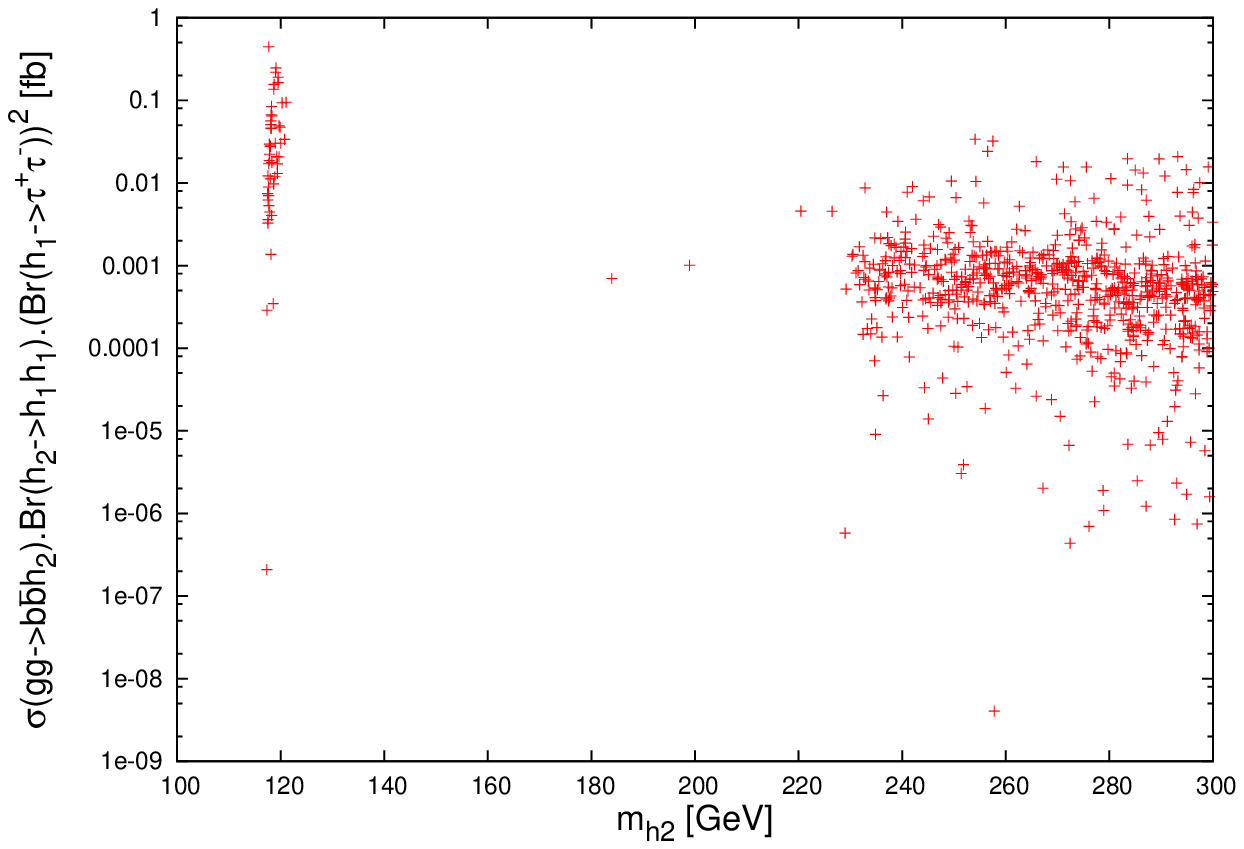}

 \end{tabular}
   \label{fig:sigma-scan5}
\caption{The signal rates for $\sigma(pp\to b\bar b {h_2})~{\rm BR}(h_2\to h_1h_1)$ 
times BR($h_1h_1\to b\bar bb\bar b$), BR($h_1h_1\to b\bar b\tau^+\tau^-$)
and BR($h_1h_1\to \tau^+\tau^-\tau^+\tau^-$)   
 as functions of $m_{h_2}$.}

\end{figure}

\begin{figure}
 \centering\begin{tabular}{cc}
  \includegraphics[scale=0.60]{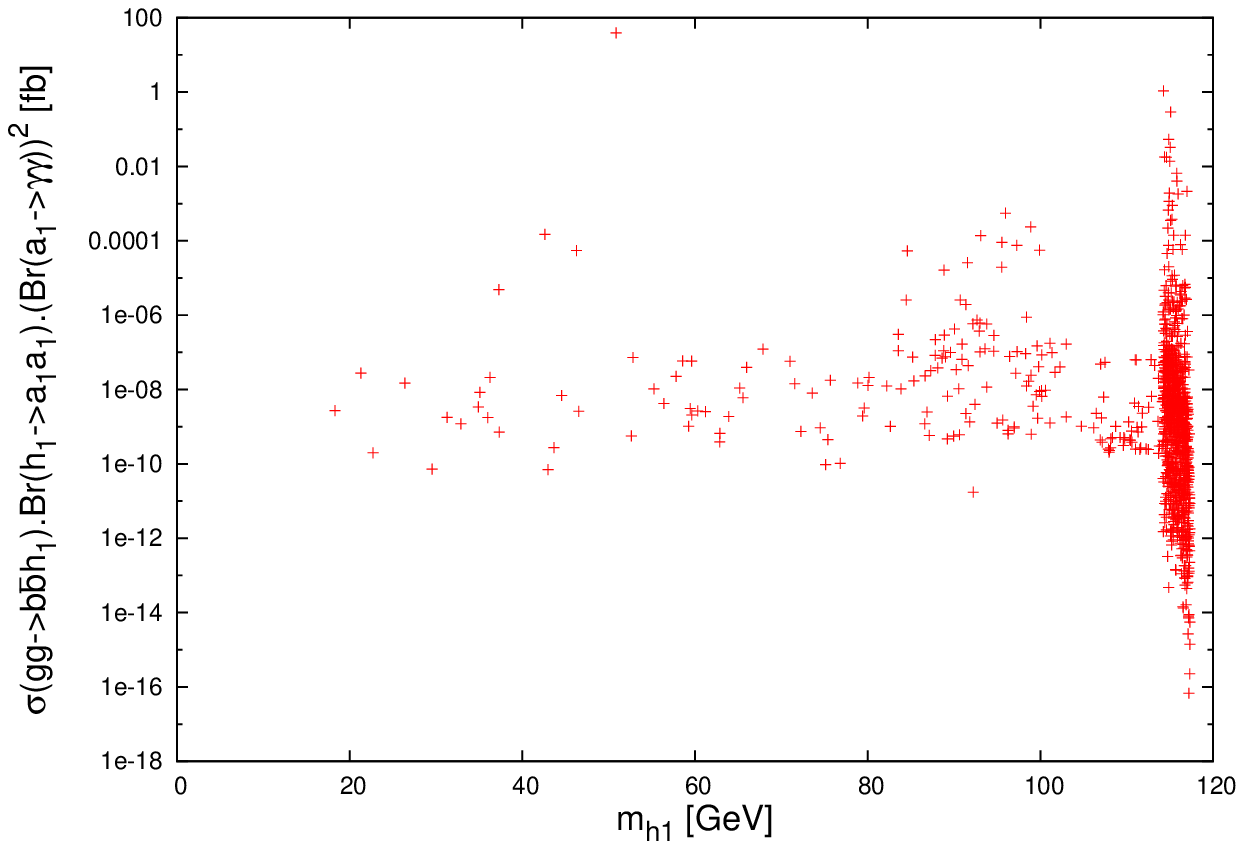}&\includegraphics[scale=0.60]{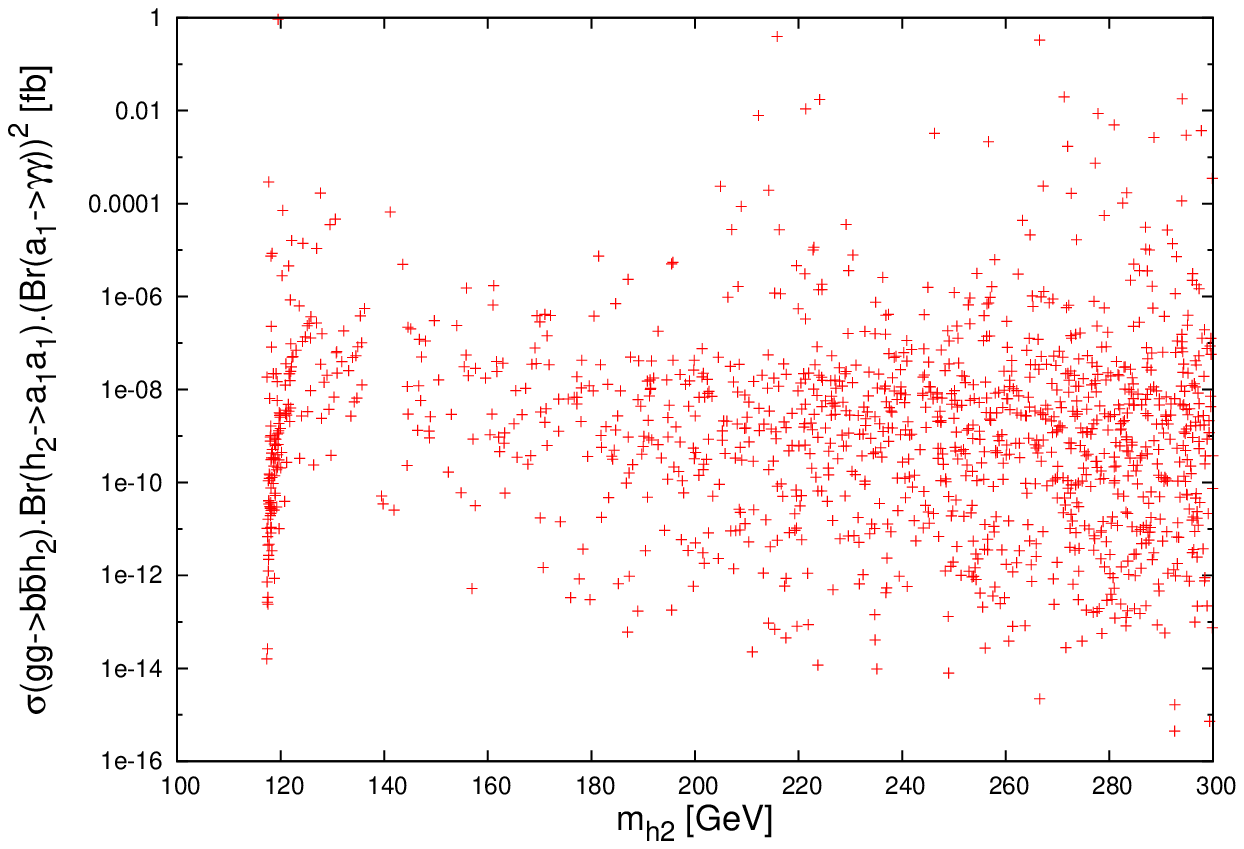}\\
\includegraphics[scale=0.60]{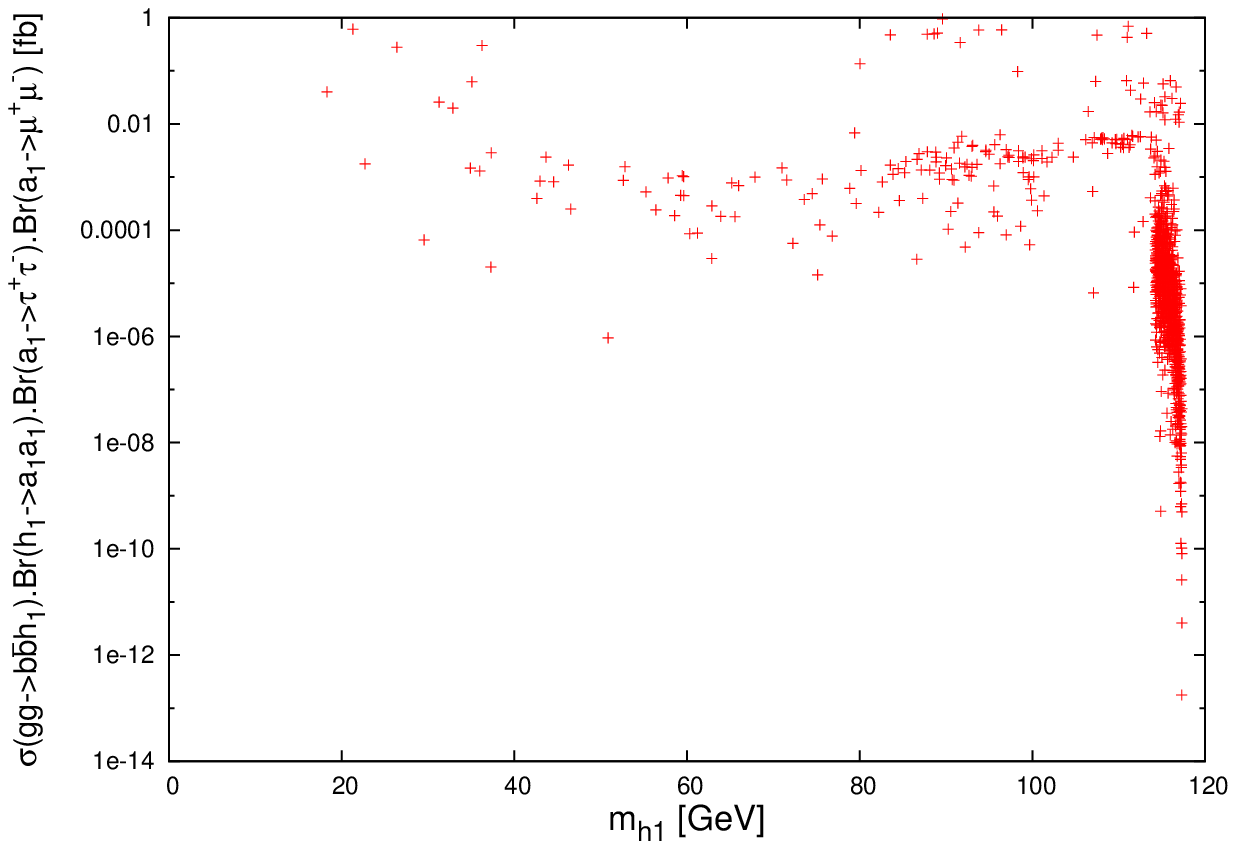}&\includegraphics[scale=0.60]{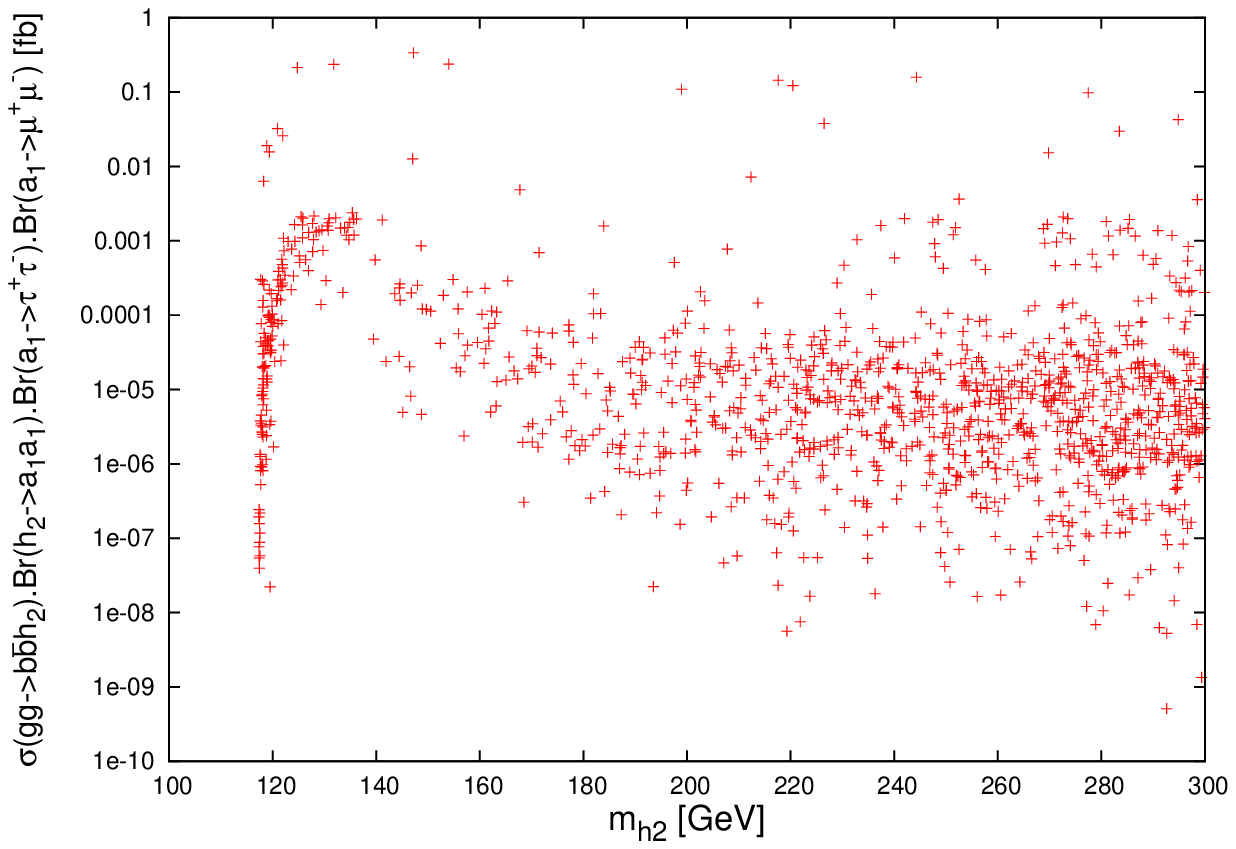}
       
 \end{tabular}
   \label{fig:sigma-scan6}
\caption{The signal rates for $\sigma(pp\to b\bar b {h_1})~{\rm BR}(h_1\to a_1a_1)$ and 
$\sigma(pp\to b\bar b {h_2})~{\rm BR}(h_2\to a_1a_1)$ times (BR($a_1\to \gamma\gamma))^2$ and times
BR$(a_1\to \tau^+\tau^-)$ BR$(a_1\to \mu^+\mu^-)$  
 as functions of $m_{h_1}$ and $m_{h_2}$.}

\end{figure}

\begin{figure}
 \centering\begin{tabular}{cc}
  \includegraphics[scale=0.60, angle=90]{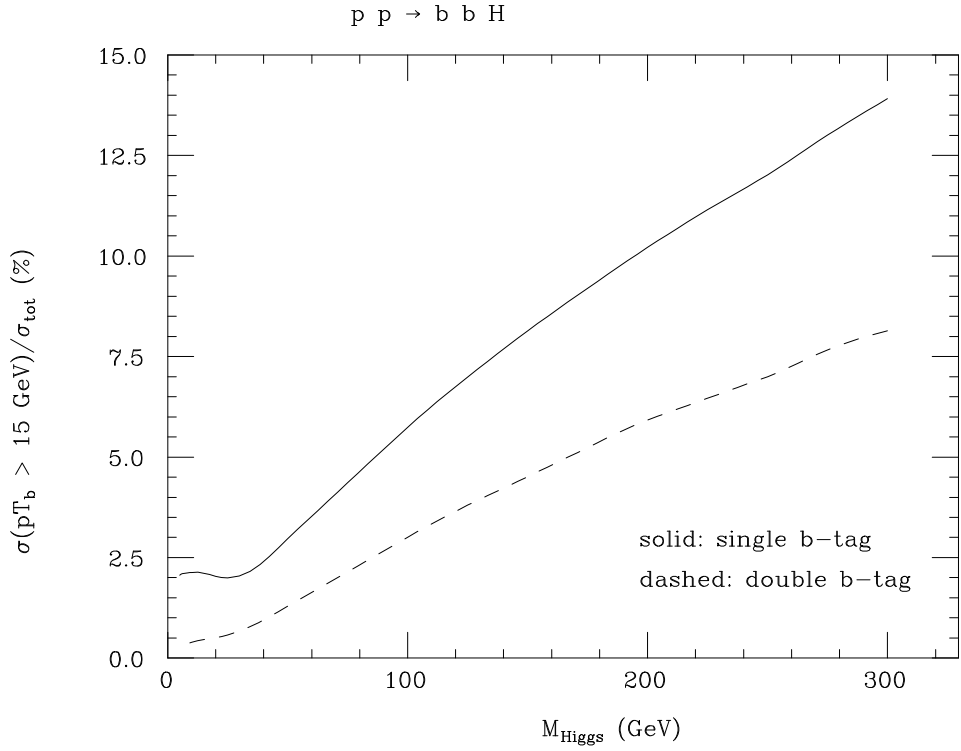}
     
 \end{tabular}
   \label{fig:pTb}
\caption{The efficiency to tag one or two `prompt' $b$-quarks in the final state, given as percent 
ratio of the production cross section for $pp\to b\bar b$ Higgs (where Higgs can equally 
refer to an $h_1$ or $h_2$ state) after
 a the cut $p_{T_b}>15$ GeV over the total one (also including the $b$-tagging performances, $\varepsilon_b$\
and $\varepsilon_b^2$, respectively), as function of the Higgs boson mass. The distributions
have been produced at parton level by using CalcHEP. Herein we use
$\varepsilon_b=60\%$.}

\end{figure}

\end{document}